\documentclass[aps,prd,11pt,nofootinbib,tightenlines,preprintnumbers,superscriptaddress,notitlepage]{revtex4-2}
\usepackage[a4paper,left=2.5cm,right=2.5cm, top=2.0cm,bottom=2.0cm]{geometry}
\usepackage[english]{babel}
\usepackage[applemac]{inputenc}
\usepackage[colorlinks,citecolor=blue,linktoc=all,linkcolor=black,urlcolor=black]{hyperref}
\usepackage{amsfonts,amsmath,amssymb}
\usepackage{graphicx}
\usepackage{booktabs, multirow}
\usepackage[detect-all]{siunitx}
\usepackage{hyperxmp}
\usepackage{color}
    
\newcommand{\Fig}[1]{Fig.~\ref{#1}}
\newcommand{\Eq}[1]{Eq.~(\ref{#1})}
\newcommand{\Section}[1]{Section~\ref{#1}}
\newcommand{\Table}[1]{Table~\ref{#1}}

\newcommand{\dof}{\mathrm{d.o.f.}}
\newcommand{\fm}{\mathrm{fm}}
\newcommand{\eV}{\mathrm{eV}}
\newcommand{\keV}{\mathrm{keV}}

\newcommand{\GeV}{\mathrm{GeV}}
	
\newcommand{\FF}{{\cal F}_{P\gamma^*\gamma^*}}
\newcommand{\FFpi}{{\cal F}_{\pi^0\gamma^*\gamma^*}}
\newcommand{\etap}{\eta^{\prime}}
\newcommand{\ahlbl}{a_\mu^{\rm hlbl}}
\newcommand{\xv}{\vec{x}}

\newcommand{\zv}{\vec{z}}
\newcommand{\pv}{\vec{p}}
\newcommand{\qv}{\vec{q}}
\newcommand{\qva}{\vec{q}_1}
\newcommand{\chib}{\overline{\chi}}
\newcommand{\Tr}{\mathrm{Tr}}
\newcommand{\amu}{a_\mu^{\mathrm{HLbL}; P}}
\newcommand{\amupi}{a_\mu^{\mathrm{HLbL}; \pi^0}}
\newcommand{\amueta}{a_\mu^{\mathrm{HLbL}; \eta}}
\newcommand{\amuetap}{a_\mu^{\mathrm{HLbL}; \etap}}
\newcommand{\alphaQED}{\alpha}

\makeatother

\begin{document}
%
\title{Lattice calculation of the $\pi^0$, $\eta$ and $\etap$ transition form factors and the hadronic light-by-light contribution to the muon $g-2$}
\author{Antoine G\'erardin}
\email{antoine.gerardin@cpt.univ-mrs.fr}
\affiliation{Aix-Marseille Universit\'e, Universit\'e de Toulon, CNRS, CPT, Marseille, France}
\author{Willem E. A. Verplanke}
\email{willem.verplanke@cpt.univ-mrs.fr}
\affiliation{Aix-Marseille Universit\'e, Universit\'e de Toulon, CNRS, CPT, Marseille, France}
\author{Gen Wang}
\affiliation{Aix-Marseille Universit\'e, Universit\'e de Toulon, CNRS, CPT, Marseille, France}
\author{Zoltan Fodor} 
\affiliation{Department of Physics, University of Wuppertal, D-42119 Wuppertal, Germany}
\affiliation{J\"ulich Supercomputing Centre, Forschungszentrum J\"ulich, D-52428 J\"ulich, Germany}
\affiliation{Institute for Theoretical Physics, E\"otv\"os University, H-1117 Budapest, Hungary}
\affiliation{Department of Physics, Pennsylvania State University, University Park, PA 16802, USA}
\author{Jana N. Guenther} 
\affiliation{Department of Physics, University of Wuppertal, D-42119 Wuppertal, Germany}
\author{Laurent Lellouch} 
\affiliation{Aix-Marseille Universit\'e, Universit\'e de Toulon, CNRS, CPT, Marseille, France}
\author{Kalman K. Szabo}
\affiliation{Department of Physics, University of Wuppertal, D-42119 Wuppertal, Germany}
\affiliation{J\"ulich Supercomputing Centre, Forschungszentrum J\"ulich, D-52428 J\"ulich, Germany}
\author{Lukas Varnhorst}
\affiliation{Department of Physics, University of Wuppertal, D-42119 Wuppertal, Germany}

\begin{abstract}
In this paper we present a first ab initio calculation of the $\pi^0$, $\eta$ and $\etap$ transition form factors performed with physical light-quark masses. We provide a complete parametrization of the form factors that includes both single and double-virtual kinematics. Our results are compared with experimental measurements of the form factors in the spacelike region and with the measured two-photon decay widths. 
In a second step, our parametrizations of the transition form factors are used to compute the dominant pseudoscalar-pole contributions to the hadronic light-by-light scattering in the muon $g-2$. Our final result reads $a_{\mu}^{\rm hlbl, ps-pole} =  (85.1 \pm 5.2) \times 10^{-11}$. Although the pion-pole is dominant, we confirm that, together, the $\eta$ and $\etap$ provide roughly half of its contribution.
\end{abstract}

\maketitle

\vspace{-0.45cm}
\section{Introduction \label{sec:introduction} }

\begin{figure}[t]
	\includegraphics*[width=0.85\linewidth]{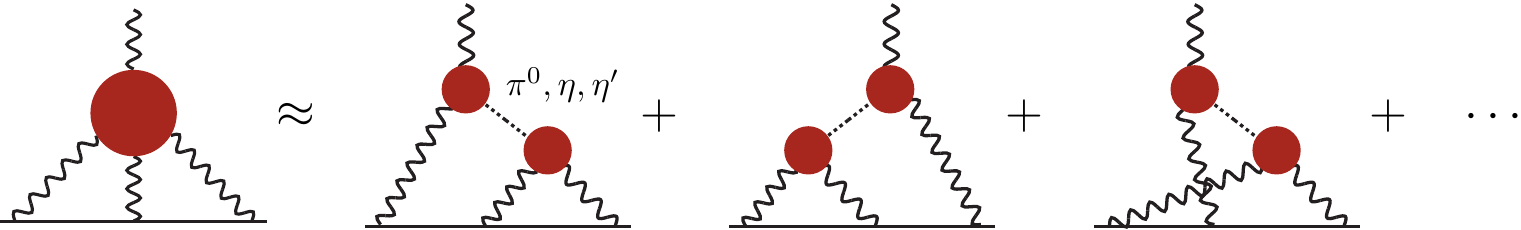}
	\caption{Pseudoscalar-pole contribution to the hadronic light-by-light scattering diagram. The blob on the left of the equality represents the full HLbL four-point function. The blobs on the right-hand side represent the transition form factors of the light pseudoscalar mesons. The solid, dashed and wavy lines represent the muon, the intermediate pseudoscalar mesons, and the photons, respectively.}	
	\label{fig:pspoles}
\end{figure}

The combined experimental result for the anomalous magnetic moment of the muon, from the Fermilab collaboration~\cite{Muong-2:2021ojo} and the E821 experiment at Brookhaven~\cite{Muong-2:2006rrc}, has reached a precision of 0.35 ppm. It translates to a 4.2$\sigma$ tension with the theory average, as published by the $g-2$ theory initiative~\cite{Aoyama:2020ynm}.
The theory error is completely dominated by effects of the strong interaction which can be separated into two distinct contributions: the hadronic vacuum polarization (HVP) which enters at order $\alphaQED^2$ and the hadronic light-by-light scattering (HLbL) at order $\alphaQED^3$. The experimental uncertainties~\cite{Muong-2:2015xgu,Abe:2019thb} are expected to be reduced by a further factor of 3 by the Fermilab experiment~\cite{Muong-2:2015xgu} and a new experiment at J-PARC~\cite{Abe:2019thb} is under construction.
To fully benefit from these two new experiments, a precision of 0.2\% for the HVP and below 10\% for the HLbL is required.

Over the past few years, lattice QCD has made significant progress toward reducing the uncertainties on the HVP contribution~\cite{DellaMorte:2017dyu, FermilabLattice:2017wgj, Budapest-Marseille-Wuppertal:2017okr, RBC:2018dos, Giusti:2019xct, Shintani:2019wai, FermilabLattice:2019ugu, Gerardin:2019rua, Aubin:2019usy, Giusti:2019hkz, Borsanyi:2020mff, Lehner:2020crt, Giusti:2021dvd,  Wang:2022lkq, Aubin:2022hgm} and the Budapest-Marseille-Wuppertal collaboration published the first subpercent calculation~\cite{Borsanyi:2020mff}. Interestingly, this result is in slight tension, at the level of 2.1$\sigma$, with the data-driven estimate~\cite{Davier:2017zfy,Keshavarzi:2018mgv,Colangelo:2018mtw,Hoferichter:2019mqg,Davier:2019can,Keshavarzi:2019abf} and tends to reduce the discrepancy between the theory estimate and the world-average measurement to the level of 1.5$\sigma$. Significant attention has been devoted to clarify this discrepancy~\cite{Colangelo:2020lcg,Crivellin:2020zul,Keshavarzi:2020bfy,Malaescu:2020zuc}. 
The situation for the HLbL contribution is different. Two groups have presented direct lattice calculations where the hadronic correlation function of four electromagnetic currents is computed on the lattice~\cite{Blum:2014oka, Blum:2015gfa, Blum:2016lnc, Blum:2019ugy, Blum:2023vlm, Green:2015sra, Green:2015mva, Gerardin:2017ryf, Chao:2020kwq, Chao:2021tvp, Asmussen:2022oql,Chao:2022xzg}. 
This correlator, computed in position space, is weighted by a kernel function that represents the QED part of the diagram (see \Fig{fig:pspoles}). 
Beside the use of different fermion formulations, the main difference between the two groups lies in the treatment of the weight function that differ between the Mainz group~\cite{Asmussen:2022oql} and the RBC/UKQCD collaboration~\cite{Blum:2023vlm}. 
Both lattice estimates are compatible with each other and with the data-driven result based on the dispersive framework developed in~~\cite{Colangelo:2014dfa, Colangelo:2014pva, Colangelo:2015ama, Colangelo:2017qdm, Colangelo:2017fiz}. However, the current precision of this comparison is at the level of 15\% and more stringent tests require improvement in both frameworks. 

On the lattice, the main challenge consists in reducing the statistical error that increases rapidly at long distances and to control possibly large finite-volume corrections. Fortunately, both sources of uncertainty can be strongly suppressed if one knows the pion transition form factor (TFF) on the same set of gauge ensembles~\cite{Chao:2022zqs}. The knowledge of the $\eta$ and $\etap$ TFFs would allow us to test the saturation of the signal at even shorter distances. 

An alternative to the direct lattice calculation, is the data driven approach~\cite{Colangelo:2014dfa, Colangelo:2014pva, Colangelo:2015ama, Colangelo:2017qdm, Colangelo:2017fiz}. It is based on a dispersive framework, similar to the one used for the HVP, but it is much more complicated due to the complex analytical structure of the four-point function involved in the HLbL diagram.  It is expected that, among all possible intermediate states, the largest contributions are given by a handful of states: the light pseudoscalar mesons ($\pi^0$,  $\eta$ and $\etap$). In this dispersive framework, the contribution from the pseudoscalar states can be unambiguously computed once the TFFs, describing the interaction of the mesons with two virtual photons, are known (see \Fig{fig:pspoles}).

Although the form factors are, in principle, accessible through experiments, their measurement in the whole kinematic range relevant for the $g-2$ is challenging. Only the single-virtual form factor has been measured by the BaBar~\cite{BaBar:2009rrj}, CELLO~\cite{CELLO:1990klc}, CLEO~\cite{CLEO:1997fho} and Belle~\cite{Belle:2012wwz} collaborations for spacelike virtualities above 0.6~$\GeV^2$. Recently, the BES III collaboration has presented preliminary results down to $0.3~\GeV^2$. 
The single-virtual $\eta$ and $\etap$ form factors have been measured in the space-like region by CELLO~\cite{CELLO:1990klc}, CLEO~\cite{CLEO:1997fho},  BaBar~\cite{BaBar:2011nrp} and the L3 experiment at LEP~\cite{L3:1997ocz}, covering the region from $0.3~\GeV^2$ up to $40~\GeV^2$. The BaBar collaboration has published first results for the double-virtual TFF associated with the $\etap$ meson, but only for virtualities larger than $Q^2 \approx 6.5~\GeV^2$~\cite{BaBar:2018zpn}.
When both photons are real, the form factors are related to the two-photon decay width. The latter has been measured for all three mesons~\cite{ParticleDataGroup:2018ovx,PrimEx-II:2020jwd} and is an important constraint to the  low-energy behavior of the TFFs.

In~\cite{Gerardin:2016cqj,Gerardin:2019vio} it has been shown that a precise lattice calculation of the pion TFF in the kinematical range relevant for the $g-2$ is possible, leading to $\amupi = (59.7 \pm 3.6) \times 10^{-11}$. This result is in good agreement with the data-driven estimate~\cite{Hoferichter:2018kwz,Hoferichter:2018dmo}. 
Using similar techniques, the $\eta$-pole contribution has been computed by the ETM collaboration, but at a single lattice spacing~\cite{Alexandrou:2022qyf}.
In this work, we provide a new estimate of the pion-pole contribution, using a different fermionic action as compared to~\cite{Gerardin:2019vio}, and we extend the calculation to the $\eta$ and $\etap$ mesons.  We provide parametrizations of the form factors at the physical point that are then used to estimate the light pseudoscalar-pole contributions to the HLbL diagram. 
The dedicated spectroscopy analysis of the $\eta$ and $\etap$ mesons using staggered quarks, on the same set of ensembles, is presented in~\cite{spectro}.

This paper is organized as follows. In \Section{sec:methodology} we describe our strategy to extract the TFFs in Euclidean spacetime using lattice QCD simulations. In \Section{sec:sim} we present our lattice setup and explain how the relevant correlation functions are computed with staggered quarks. Then, in \Section{sec:analysis}, we present our analysis and we discuss the main sources of systematic uncertainties. 
Finally, in \Section{sec:results}, we present our results for all three TFFs and we discuss phenomenological applications. 
Tables with the coefficients of the TFF parametrizations in the continuum limit, and the associated covariance matrices, are provided in Appendix~\ref{app:tables}. In Appendix~\ref{sec:discV}, we compare different estimators to compute single vector current insertions in a quark loop.

\section{Methodology \label{sec:methodology} }

\subsection{Form factors}

In this section, we follow the notations introduced in \cite{Gerardin:2016cqj,Gerardin:2019vio}.  The TFF describing the interaction between a pseudoscalar meson $P=\pi^0, \eta, \etap$, with momentum $p$, and two off-shell photons, with momenta $q_1$ and $q_2$, such that $p=q_1+q_2$, is defined via the following matrix element
\begin{equation} 
M_{\mu\nu}(p,q_1)  = i \int \mathrm{d}^4 x \, e^{i q_1 \cdot x} \, \langle \Omega | T \{ J_{\mu}(x) J_{\nu}(0) \} | P(p) \rangle =
\epsilon_{\mu\nu\alpha\beta} \, q_1^{\alpha} \, q_2^{\beta} \, \FF(q_1^2, q_2^2) \,.
\end{equation}
In this equation, $J_{\mu}$ is the hadronic component of the electromagnetic current and $\epsilon_{\mu\nu\alpha\beta}$ is the fully antisymmetric tensor with $\epsilon^{0123} = 1$. 
We have implicitly treated the meson as an asymptotic state. 
This is true for the pion but also for the $\eta$ meson in the isospin limit of QCD (the decays $\eta \to \pi^+ \pi^- \pi^0$ and $\eta \to 3\pi^0$ break isospin symmetry). 
On the other hand, the $\etap$ meson can decay via the strong interaction, even in the isospin limit, and its main decay modes are $\etap \to \eta \pi^+ \pi^-$ and $\etap \to \eta \pi^0 \pi^0$.
However, the smallness of the decay width justifies the narrow-width approximation in which the meson is treated as stable particle. 
Below the threshold for hadron production in the vector channel,\footnote{In the isovector case, the threshold $s_0$ is given by $4m_\pi^2$.} the TFF is obtained in Euclidean spacetime after analytical continuation~\cite{Ji:2001wha,Ji:2001nf}. 
Using the superscript E, the Euclidean matrix element reads
\begin{equation}
M_{\mu\nu}  =  (i^{n_0}) M_{\mu\nu}^{\rm E}, \quad 
M_{\mu\nu}^{\rm E}  \equiv  - \int_{-\infty}^{+\infty} \mathrm{d} \tau \, e^{\omega_1 \tau}  \int \mathrm{d}^3 z \, e^{-i \vec{q}_1 \cdot \vec{z}} \, 
\langle 0 | T \left\{ J_{\mu}(\vec{z}, \tau) J_{\nu}(\vec{0}, 0) \right\} | P(p) \rangle  \,,
\label{eq:def}
\end{equation}
where $n_0$ denotes the number of temporal indices carried by the two vector currents. Different virtualities can be reached by tuning the real, free parameter $\omega_1$ such that $q_1 = (\omega_1, \vec{q}_1)$ and $q_2 = p - q_1$.
It is convenient to write \Eq{eq:def} as an integral over $\tau$, the time separation between the two vector currents\footnote{Note that our conventions slightly differ from \cite{Gerardin:2019vio}.}
\begin{equation}
 M_{\mu\nu}^{\rm E} = \int_{-\infty}^{\infty} \, \mathrm{d}\tau \, \widetilde{A}^{(P)}_{\mu\nu}(\tau) \, e^{\omega_1 \tau}  \,.
\label{eq:Mlat}
\end{equation}

Since we are working in the isospin limit with equal-mass $u$ and $d$ quarks, the pion does not mix with the $\eta$ and $\etap$ mesons and an interpolating operator for this meson is given by the pseudoscalar density $P_{3}(x)$ below.
If we further assume SU(3) flavor symmetry then the $\eta$ and $\etap$ mesons can be identified with the octet $\eta_8$ and singlet $\eta_0$ pseudoscalar mesons. Those states can be studied using the densities $P_{8}(x)$ and $P_{0}(x)$, respectively. We have
\begin{subequations}
\begin{align}
	P_{3}(x) &= \frac{1}{\sqrt{2}} \left(\overline{u}\gamma_5 u(x) - \overline{d}\gamma_5 d(x)\right), \\
	P_{8}(x) &= \frac{1}{\sqrt{6}}\left(\overline{u}\gamma_5 u(x) + \overline{d}\gamma_5 d(x) - 2\overline{s}\gamma_5 s(x)\right),\\
	P_{0}(x) &= \frac{1}{\sqrt{3}} \left(\overline{u}\gamma_5 u(x) + \overline{d}\gamma_5 d(x) + \overline{s}\gamma_5 s(x)\right).
\end{align}
\label{eq:op}
\end{subequations}
Away from the SU(3) flavor limit, the $\eta_8$ and $\eta_0$ states mix to give the physical $\eta$ and $\etap$ mesons that have nonvanishing overlaps with both $P_{8}$ and $P_{0}$ . We now explain how this effect is taken into account in this work. 

To extract the matrix element in \Eq{eq:def}, we consider the following three-point correlation functions 
\begin{equation}
C^{(i)}_{\mu\nu}(\tau,t_{P}) = \sum_{\xv,\zv} \langle J_{\mu}(\zv,t_i) J_{\nu}(\vec{0},t_f) \mathcal{O}_i(\xv,t_0) \rangle \, e^{i \pv \cdot \xv}  e^{-i \qva \cdot \zv} \,,
\label{eq:3pt}
\end{equation}
where $t_{P}={\rm min}(t_f-t_0,t_i-t_0)$ is the minimal time separation between the pseudoscalar interpolating operator and either of the two vector currents and $\tau = t_i-t_f$. The lattice operators $\mathcal{O}_i$ have the same quantum numbers as the pseudoscalar densities $P_i$. The precise definition of those operators will be given in \Section{sec:stagg}. 
For the pion, neglecting excited states and wrap around contributions, we obtain the asymptotic behavior
\begin{align}
C^{(3)}_{\mu\nu}(\tau,t_{P}) =& \sum_{\zv} \langle 0 | J_{\mu}(\zv, \tau) J_{\nu}(\vec{0},0) | \pi(\pv) \rangle  e^{-i \qva \cdot \zv} \times  \frac{1}{2E_\pi} \langle \pi(\pv) | \mathcal{O}_3 | 0 \rangle e^{ E_\pi (t_0 - t_f)} 
\end{align}
where $E_\pi$ is the energy of the pion and $Z_{\pi}=\langle \pi |\mathcal{O}_3| 0\rangle$ is the overlap factor of the pseudoscalar operator with the pion state. It leads to
\begin{equation}
\widetilde{A}^{(\pi)}_{\mu\nu}(\tau) =  \lim_{t_{P} \rightarrow + \infty} \widetilde{A}_{\mu\nu}^{(\pi); {\rm eff}}(\tau,t_P)\,, \quad \widetilde{A}_{\mu\nu}^{(\pi); {\rm eff}}(\tau,t_P) = \frac{2E_{\pi}}{Z_{\pi}} e^{E_\pi (t_f-t_0)} C^{(3)}_{\mu\nu}(\tau,t_{P}) \,.
\label{eq:Amunu}
\end{equation}

A similar expression could be used for the $\eta$ meson, looking at the correlator $C^{(8)}_{\mu\nu}(\tau,t_P)$. Indeed, the correlation function $C^{(8)}_{\mu\nu}(\tau,t_P)$ is expected to be dominated by the $\eta$ meson, with little contribution from the heavier $\etap$ that can be seen as an excited state. This is the procedure followed in~\cite{Alexandrou:2022qyf}. However, this method has some disadvantages. First, the long time asymptotic is reached at large values of $t_{P}$ where the signal-to-noise ratio deteriorates rapidly. Second, this method is not applicable to the $\etap$ state. A comparison of two approaches for the $\eta$ meson is presented in~\Section{sec:topt}.

Instead, the $\eta$ and $\etap$ TFFs are extracted from a $2 \times 2$ matrix of three-point functions. Neglecting higher excited states, the spectral decomposition for $i=0,8$ now reads 
\begin{align}
\nonumber C^{(i)}_{\mu\nu}(\tau,t_{P}) =& \sum_{\zv} \langle 0 | J_{\mu}(\zv,\tau) J_{\nu}(\vec{0},0) | \eta(\pv) \rangle  e^{-i \qva \cdot \zv} \times  \frac{1}{2E_\eta} \langle \eta(\pv) | \mathcal{O}_i | 0 \rangle e^{E_\eta (t_0 - t_f)}  \\
+ &\sum_{\zv} \langle 0 | J_{\mu}(\zv,\tau) J_{\nu}(\vec{0}, 0) | \etap(\pv) \rangle e^{-i \qva \cdot \zv} \times  \frac{1}{2E_{\etap}} \langle \etap(\pv) | \mathcal{O}_i | 0 \rangle e^{E_\etap (t_0 - t_f)} \,. \label{eq:dec}
\end{align}
Or, in a more compact matrix notation
\begin{equation}
\begin{pmatrix}
C_{\mu\nu}^{(8)} \\[1.5mm]
C_{\mu\nu}^{(0)}
\end{pmatrix} =
\begin{pmatrix}
T^{(8)}_{\eta} & T^{(8)}_{\etap} \\
T^{(0)}_{\eta} & T^{(0)}_{\etap} 
\end{pmatrix} 
\begin{pmatrix}
\widetilde{A}^{(\eta)}_{\mu\nu} \\[1.3mm]
\widetilde{A}^{(\etap)}_{\mu\nu}
\end{pmatrix} \,,
\label{eq:matrix}
\end{equation}
with
\begin{equation}
T_n^{(i)} = \frac{Z_n^{(i)}}{2E_n} e^{-E_n (t_f-t_0) } \,, \quad  \widetilde{A}^{(n)}_{\mu\nu}(\tau) = \sum_{\zv} \langle 0 | J_{\mu}(\zv,\tau) J_{\nu}(\vec{0},0) | n(\pv) \rangle \, e^{-i \qva \cdot \zv} \,.
\end{equation}
The four overlap factors are defined by $Z_n^{(i)} = \langle n | \mathcal{O}_i | 0 \rangle$. Inverting \Eq{eq:matrix}, one obtains the $\eta$ and $\etap$ TFFs in terms of the correlators computed on the lattice
\begin{subequations}
\begin{align}
\widetilde{A}^{(\eta)}_{\mu\nu} &=  \cos^2 \phi_I \ \frac{C_{\mu\nu}^{(8)}}{T_{\eta}^{(8)}} +  \sin^2 \phi_I \ \frac{C_{\mu\nu}^{(0)} }{T_{\eta}^{(0)}}  \\
\widetilde{A}^{(\etap)}_{\mu\nu} &= \sin^2 \phi_I \ \frac{C_{\mu\nu}^{(8)}}{T_{\etap}^{(8)}}   + \cos^2 \phi_I \ \frac{C_{\mu\nu}^{(0)}}{T_{\etap}^{(0)}}  
\end{align}
\label{eq:improved}
\end{subequations}
where the mixing angle in the isospin basis, $\phi_I$, is given by $\tan^2\phi_I = - ( Z^{(8)}_{\etap} Z^{(0)}_{\eta} ) / ( Z^{(8)}_{\eta} Z^{(0)}_{\etap} )$.

In practice, the masses and overlap factors $Z_n^{(i)}$ can be extracted by solving the generalized eigenvalue problem (GEVP) using a matrix of pseudoscalar two-point correlation functions 
\begin{equation}
C^{\rm 2pt}(t) v_n(t,t_0) = \lambda_n(t,t_0) C^{\rm 2pt}(t_0) v_n(t,t_0) \,,
\end{equation}
where $C^{\rm 2pt}_{ij}(t) = \langle \mathcal{O}_i(t) \mathcal{O}_j(0) \rangle$ with $i,j \in 0,8$ and with $t_0$ a free parameter. 
The effective masses can be extracted from the logarithmic derivative of the eigenvalues $\lambda_n$~\cite{Blossier:2009kd}
\begin{equation}
am_n^{\rm eff}(t) = \log \frac{ \lambda_n(t,t_0) }{ \lambda_n(t+a,t_0) } 
\end{equation}
that can be eventually fitted to a constant at large times. The overlap factors can be obtained from the eigenvectors $v_n$ through
\begin{equation}
Z_{n}^{{(i)}\, {\rm eff}}(t) = \sqrt{2 E_n} \ \sum_{j} C^{\rm 2pt}_{ij}(t) \, v_{nj}(t,t_0) \times \left( \frac{\lambda_n(t,t_0)}{ \lambda_n(t+a,t_0) } \right)^{t/a - t_0/(2a)} 
\end{equation}
assuming the normalization
\begin{equation}
v_m(t,t_0)^{\rm T} \, C^{\rm 2pt}(t_0) \, v_n(t,t_0) = \delta_{n,m} \,.
\end{equation}
However, we prefer to follow the strategy presented in~\cite{spectro} where a fit of the pseudoscalar two-point correlation matrix is performed. This method leads to numerically more stable results.

\subsection{Pseudoscalar-pole contribution to HLbL scattering in $(g-2)_{\mu}$}

\begin{table}[t]
\renewcommand{\arraystretch}{1.1}
\caption{Parameters of the simulations: the bare coupling $\beta = 6/g_0^2$, the lattice dimensions $L^3 \times T$, the lattice spacing $a$ and the spatial extent $L$ in physical units, the bare light and strange quark masses, the mass of the taste-singlet pion and the number of gauge configurations. The last three columns indicate which TFFs have been computed.}
\vskip 0.1in
\begin{tabular}{lcl@{\hskip 01em}c@{\hskip 02em}l@{\hskip 01em}c@{\hskip 02em}l@{\hskip 01em}l@{\hskip 01em}l@{\hskip 01em}@{\hskip 01em}l@{\hskip 01em}l@{\hskip 01em}l}
\hline
$\quad\beta\quad$ & $L^3\times T$ & $a~[\fm]$ & $L~[\fm]$ & $am_l$ & $am_s$ & $m_{\pi}^I$~[MeV]  & $\#$~cnfgs &$\pi^0$ & $\eta$ & $\etap$ \\
\hline
$3.7000$	& $24^3\times48$	& 0.1315	& 3.2 & 0.00205349 & 0.0572911	& 430	& 700	& \checkmark & \checkmark & \checkmark   \\
		& $32^3\times64$	&		& 4.2 & 0.00205349 & 0.0572911	& 		& 900	& \checkmark & \checkmark & \checkmark   \\
		& $48^3\times64$	&		& 6.3 & 0.00205349 & 0.0572911	& 		& 900	& \checkmark & \checkmark & \checkmark   \\
\hline
$3.7500$	& $56^3\times96$	& 0.1191	& 6.7 & 0.00184096 & 0.0495930	& 380	& 500	& \checkmark &  $-$ & $-$  \\
		& $56^3\times96$	&		& 6.7 & 0.00176877 & 0.0516173	& 		& 500	& \checkmark &  $-$ & $-$  \\
		& $56^3\times96$	&		& 6.7 & 0.00184096 & 0.0516173	& 		& 500  	& \checkmark &  $-$ & $-$  \\
\hline
$3.7753$	& $28^3\times56$	& 0.1116	& 3.1 & 0.00171008 & 0.0476146	& 355	& 850	& \checkmark & \checkmark & \checkmark   \\
		& $56^3\times84$	&		& 6.2 & 0.00171008 & 0.0476146	& 		& 500	& \checkmark &  $-$ & $-$  \\
		& $56^3\times84$	&		& 6.2 & 0.00171008 & 0.0485669	& 		& 500	& \checkmark &  $-$ & $-$  \\
		& $56^3\times84$	&		& 6.2 & 0.00174428 & 0.0461862	& 		& 500	& \checkmark & \checkmark & \checkmark   \\
\hline
$3.8400$	& $32^3\times64$	& 0.0952	& 3.0 & 0.00151556 & 0.0431935	& 290	& 1100	& \checkmark & \checkmark & \checkmark   \\
		& $32^3\times64$	&		& 3.0 & 0.00143       & 0.0431935	& 		& 1050	& \checkmark & \checkmark & \checkmark   \\
		& $32^3\times64$	&		& 3.0 & 0.001455     & 0.04075		& 		& 1100	& \checkmark & \checkmark & \checkmark   \\
		& $32^3\times64$	&		& 3.0 & 0.001455     & 0.03913		& 		& 1100	& \checkmark & \checkmark & \checkmark   \\
		& $64^3\times96$	&		& 6.1 & 0.00151556 & 0.0431935	& 		& 500 	& \checkmark & $-$	   & $-$  \\
		& $64^3\times96$	&		& 6.1 & 0.001455     & 0.04075		& 		& 1100	& \checkmark & $-$	   & $-$  \\
\hline
$3.9200$	& $40^3\times80$	& 0.0787	& 3.1 & 0.001207     & 0.032		& 230	& 550	& \checkmark & \checkmark & \checkmark  \\
		& $40^3\times80$	&		& 3.1 & 0.0012         & 0.0332856	& 		& 350	& \checkmark & \checkmark & \checkmark  \\
		& $80^3\times128$	&		& 6.3 & 0.001172     & 0.03244		& 		& 500 	& \checkmark & $-$	   & $-$  \\
		& $80^3\times128$	&		& 6.3 & 0.0012         & 0.0332856	& 		& 500	& \checkmark & $-$	   & $-$  \\
\hline
$4.0126$	& $48^3\times96$	& 0.0640	& 3.1 & 0.00095897 & 0.0264999	& 190	& 850	& \checkmark & \checkmark & \checkmark  \\
		& $48^3\times96$	&		& 3.1 & 0.001002     & 0.027318	& 		& 450  	& \checkmark & \checkmark & \checkmark  \\
		& $96^3\times144$	&		& 6.1 & 0.000977     & 0.0264999	& 		& 500 	& \checkmark & $-$	   & $-$  \\
		& $96^3\times144$	&		& 6.1 & 0.001002     & 0.027318	& 		& 450 	& \checkmark & $-$	   & $-$  \\
\hline
\end{tabular} 
\label{tab:ens}
\end{table}

The TFFs are the key ingredients to compute the pseudoscalar-pole contribution to the HLbL diagram in the dispersive framework introduced in~\cite{Colangelo:2014dfa,Colangelo:2014pva,Colangelo:2015ama,Colangelo:2017qdm,Colangelo:2017fiz}. 
The master equation, first derived in Ref.~\cite{Jegerlehner:2009ry}, holds exactly for the pion and the $\eta$ in the isospin limit and for the $\etap$ in the narrow-width approximation.
Starting from the two-loop integral (see \Fig{fig:pspoles}), all angular integrals, except one, are performed using the Gegenbauer polynomial technique leading to
\begin{multline}
a_\mu^{\mathrm{HLbL}; P}  =  \left( \frac{\alphaQED}{\pi} \right)^3 \int_0^\infty \!\!\!dQ_1 \!\!\int_0^\infty \!\!\!dQ_2 \!\! \int_{-1}^{1} \!\!d\beta \,   \Big( w_1(Q_1,Q_2,\beta) \, \FF(-Q_1^2, -(Q_1 + Q_2)^2) \times   \\ 
\FF(-Q_2^2,0)  + w_2(Q_1,Q_2,\beta) \, \FF(-Q_1^2, -Q_2^2) \, \FF(-(Q_1+Q_2)^2,0) \Big) \,,
\label{eq:master}
\end{multline}  
where $w_1$ and $w_2$ are two model-independent weight functions, $Q_1$ and $Q_2$ are the norms of the two Euclidean four-momentum vectors and  $\beta = \cos \theta$ is related to the angle between them, $Q_1 \cdot Q_2 = Q_1Q_2 \cos \theta$. 
The weight functions and the shape of the integrand will be further discussed in \Section{sec:g-2}, along with the presentation of the lattice data.

\section{Simulation details \label{sec:sim}} 

\subsection{Lattice ensembles \label{sec:ens}}

This work is based on a subset of gauge ensembles generated by the Budapest-Marseille-Wuppertal collaboration~\cite{Borsanyi:2020mff}. They have been generated using $N_f = 2+1+1$ dynamical staggered fermions with four steps of stout smearing. 
The rooting procedure is used to reduce the number of tastes from four to one in the sea. This is justified in the continuum limit due to the exact fourfold degeneracy~\cite{Sharpe:2006re} but this method is more controversial at finite lattice spacings where taste violation effects break the degeneracy among tastes, leading to a nonunitary theory~\cite{Creutz:2007rk,Creutz:2007yg}. 
These unphysical effects are, however, expected to vanish in the continuum limit~\cite{Sharpe:2006re,Bernard:2006zw,Kronfeld:2007ek,Golterman:2008gt,Adams:2008db}.
In particular, our recent work~\cite{spectro} suggests that staggered fermions do reproduce the $\eta^{\prime}$ mass in the continuum limit.
The bare quark masses have been tuned such that the Goldstone mesons are at nearly physical pion and kaon mass. The lattice spacing is set using the $\Omega$ baryon mass and we exploit six values of the lattice spacing in the range $[0.0640-0.1315]$~fm to extrapolate our result to the continuum limit. We also consider $L=3$, $4$ and $6$~fm boxes for finite-size effect studies. Simulations are performed in the isospin limit where $m_u = m_d \equiv m_\ell$. More details about those ensembles can be found in~\cite{Borsanyi:2020mff} and the main properties relevant to this work are summarized in \Table{tab:ens}.

\subsection{Staggered correlation functions \label{sec:stagg}}

The extraction of the TFFs requires the calculation of the three-point functions in \Eq{eq:3pt} and the corresponding pseudoscalar two-point functions to extract the pseudoscalar masses and overlap factors that appear in \Eq{eq:improved}. We now describe the construction of those correlators.

In the staggered formalism, the operators are chosen such that the overall correlation function is a taste singlet. 
Since we are interested in the extraction of the $\eta^{(\prime)}$ TFFs, both the two- and three-point correlation functions involve Wick contractions with quark-disconnected contributions for the pseudoscalar density. Such contributions require taste-singlet pseudoscalar operators.
Using the notations of~\cite{Follana:2006rc}, where operators are given in the staggered spin-taste basis as $\Gamma_S \otimes \Gamma_T$, the operator would correspond to a $\gamma_5 \otimes 1$ operator. 
In~\cite{spectro} we consider two operators~\cite{Altmeyer:1992dd}: a time-local 3-link operator and a time nonlocal 4-link operator where the operator has support on two timeslices. 
In this work, we use the 4-link operator that has the advantage to highly suppress the parity partner state such that no oscillation in time is observed in the analysis of the two-point correlation functions. 
It makes the subsequent analysis of the two and three point functions much simpler.

For the vector current, we also use a taste singlet operator. More precisely, we use the conserved one-link operators
\begin{equation}
J_{\mu}(x) = -\frac{1}{2} \eta_{\mu}(x) \left[ \chib(x+a\hat{\mu}) U_{\mu}^{\dag}(x) \chi(x) + \chib(x) U_{\mu}(x) \chi(x+a\hat{\mu}) \right] \,,
\label{eq:V}
\end{equation}
where $\chi$ is a single component staggered fermion field and $\eta_{\mu}$ is a phase factor (we use the convention of \cite{Follana:2006rc}, also used in~\cite{spectro}). The gauge links $U_{\mu}(x)$ ensure that correlation functions are gauge invariant. 
In \Eq{eq:V}, the current is defined for a single flavor and the electromagnetic charge factors are not included.
Our choice is motivated by the presence of disconnected contributions that involve a single vector loop and also require a taste-singlet operator. Moreover, this current does not require multiplicative renormalization. We note that this operator was already used in~\cite{Borsanyi:2020mff} to compute the LO-HVP contribution.

Among the 16 tastes, the taste-singlet pion is the heaviest. Taste splitting is expected to decrease as O($a^2$) as we approach the continuum limit~\cite{Lepage:1997id,Lepage:1998vj}.
Between our coarsest and finest lattice spacings the taste-singlet pion mass decreases from about 430 to 190~MeV while the goldstone pion $\gamma_5 \otimes \gamma_5$ is kept at the physical pion mass: a strong dependence on the lattice spacing is possible.

\subsection{Computational strategy}

\begin{figure}[t]
	\includegraphics*[width=0.99\linewidth]{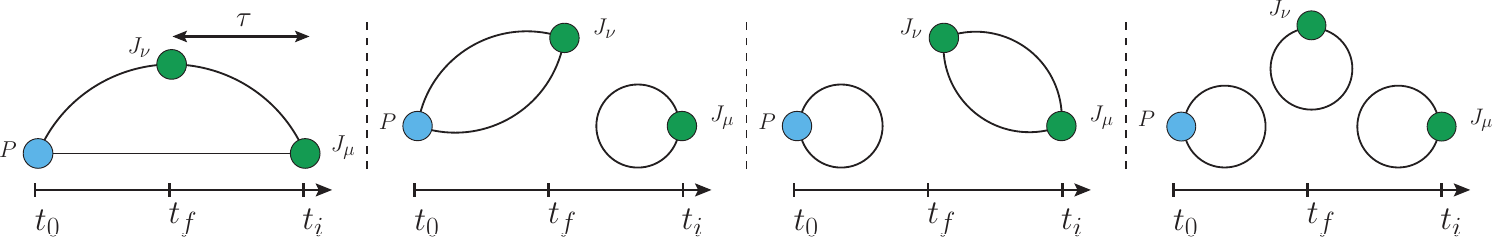} 
	\caption{The four different Wick contraction topologies that are present in the three-point correlation function. Some topologies contains several diagrams that are not shown for brevity. The blue and green blobs represent a pseudoscalar or a vector current insertion respectively. Only the first two contractions contribute to the pion TFF.}	
	\label{fig:wick}
\end{figure}

The three-point correlation function in \Eq{eq:3pt} receives potential contributions from four Wick contractions, shown in \Fig{fig:wick}. In this section, the pseudoscalar and vector currents are defined for a single flavor and the correct charges factors associated with the operators in \Eq{eq:op} need to be included when building the $\pi^0$, $\eta$ and $\etap$ correlators.

The connected contribution to the three-point function is computed using similar techniques to the one used in~\cite{Gerardin:2016cqj,Gerardin:2019vio}. First, a point source is created at a random position on the timeslice $t_f$. For each value of the meson momentum $\vec{p}$, the pseudoscalar operator is applied to the solution vector, restricted to the timeslice $t_0$, and then used as a source for a sequential inversion. 
The correlation function is finally obtained by contracting the original and the sequential solution vectors with appropriate phase factors and shifts on timeslice $t_i$. 
This method allows us to access many photon virtualities $\vec{q}_1$ and $\vec{q}_2$ at no additional cost.  Only additional pion momenta, or additional values of $t_{P}$, defined after~\Eq{eq:3pt}, require new evaluations of the quark propagator. 

The main computational challenge is to obtain a good signal for the quark-disconnected contributions. 
First, we need to compute the pseudoscalar and vector loop functions
\begin{subequations}
\label{eq:loops}
\begin{align}
L_P^{(f)}(x_0,\pv) &=  - \left(\frac{a}{L}\right)^{3} \sum_{\xv} \Tr \left[ \mathcal{O}_P \, S_f(x,x) \, \right] \, e^{ i \pv \cdot \xv} \\
L_{V;\mu}^{(f)}(x_0;\qva) &= + \frac{1}{2} \left(\frac{a}{L}\right)^{3} \sum_{ \xv } \, \eta_{\mu}(x) \, \Tr \left[ S_f(x, x+a\hat{\mu})  U_{\mu}^{\dag}(x)  + S_f(x+a\hat{\mu}, x) U_{\mu}(x) \right] \, e^{i \qva \cdot \xv } 
\end{align}
\end{subequations}
with $S_f$ the quark propagator of flavor $f=(l,s)$, $\mathcal{O}_P$ the pseudoscalar taste-singlet 4-link operator presented in~\cite{spectro} and where the trace runs over color indices. Those functions have been stored on disk for all time slices and all required values of the momenta. 
The pseudoscalar loops have been computed using a combination of low-mode averaging (LMA)~\cite{DeGrand:2004wh,Giusti:2004yp} and all-mode-averaging (AMA)~\cite{Bali:2009hu,Blum:2012uh,Shintani:2014vja} combined with the Venkataraman-Kilcup variance reduction (VKVR) trick~\cite{Gregory:2007ev} to reach the intrinsic gauge noise. More details are provided in Ref.~\cite{spectro}. For the vector loops, the VKVR trick is not applicable. Instead, since we are only interested in the light minus strange combination $L_V^{(l-s)}$ (this is the only combination that appears in the isospin limit of QCD), we adopt the split-even stochastic estimator introduced in~\cite{Giusti:2019kff} combined with LMA and AMA to reach the gauge noise (see Appendix~\ref{sec:discV} for more details).

Since we want to access many photon virtualities, the vector-vector two-point correlation functions with flavor $f$, and denoted $D^{(f)}_{VV; \mu\nu}(t_f,t_i;\pv, \qva)$ (third diagram in \Fig{fig:wick}), has been computed using point sources for all source ($t_i$) and sink ($t_f$) time slices. 
It is a noisy contribution and we found that it was optimal to use AMA with 64 sloppy inversions per exact solve for the light quark and 16 for the strange quark. Sloppy inversions are defined by a fixed number of iterations, chosen such that the bias correction is small as compared to the statistical error. In practice, we used 1, 2 or 3 exact solves per time slice depending on the ensemble. 
The corresponding three-point correlators, with all flavor combinations, are finally obtained through 
\begin{equation}
C^{(3)}_{\mathrm{P}_{f_2}-\mathrm{VV}_{f_1}; \mu\nu}(\tau,t_P; \pv,\qva) =  \frac{a}{T} \sum_{t_i} \ \langle D^{(f_1)}_{VV; \mu\nu}(t_i+\tau,t_i;\pv, \qva) L^{(f_2)}_P(t_i- t_P,\pv)   \rangle \,,
\end{equation}
where translation invariance in the time direction has been used.

The computation of the pseudoscalar-vector two-point function (second diagram in \Fig{fig:wick}) is similar, except that we use a single stochastic source per time slice, at the pseudoscalar insertion, but with support on the whole time slice. We again use AMA to reduce the cost of the analysis. Since this contribution is much smaller, we use only 16 sloppy inversions per exact solve (2 for the strange quark). In this case, the final correlators for both flavors $(f)$ in the PV loop are computed as
\begin{multline*}
C^{(3)}_{\mathrm{PV}_{f}-\mathrm{V}; \mu\nu}(\tau,t_P; \pv,\qva)   =  \frac{a}{T} \sum_{t_i} \ 
\langle L^{(l-s)}_{V;\nu}(t_i;\qva-\pv) D^{(f)}_{V_{\mu}P}(t_i+\tau,t_i-t_P;\pv, \qva) \\ + L^{(l-s)}_{V;\mu}(t_i+\tau;-\qva)  D^{(f)}_{V_{\nu}P}(t_i,t_i-t_P;\pv, \pv-\qva) \rangle \,.
\end{multline*}
The remaining correlation function (last diagram in \Fig{fig:wick}) is the purely disconnected contribution, $C^{(3)}_{\mathrm{P}_{f}-\mathrm{V}-\mathrm{V}; \mu\nu}(\tau,t_P; \pv,\qva)$. It can be easily computed from the loop functions in~\Eq{eq:loops}. Again, more information is provided in Appendix~\ref{sec:discV}.

\begin{figure}[t]
	\includegraphics*[width=0.31\linewidth]{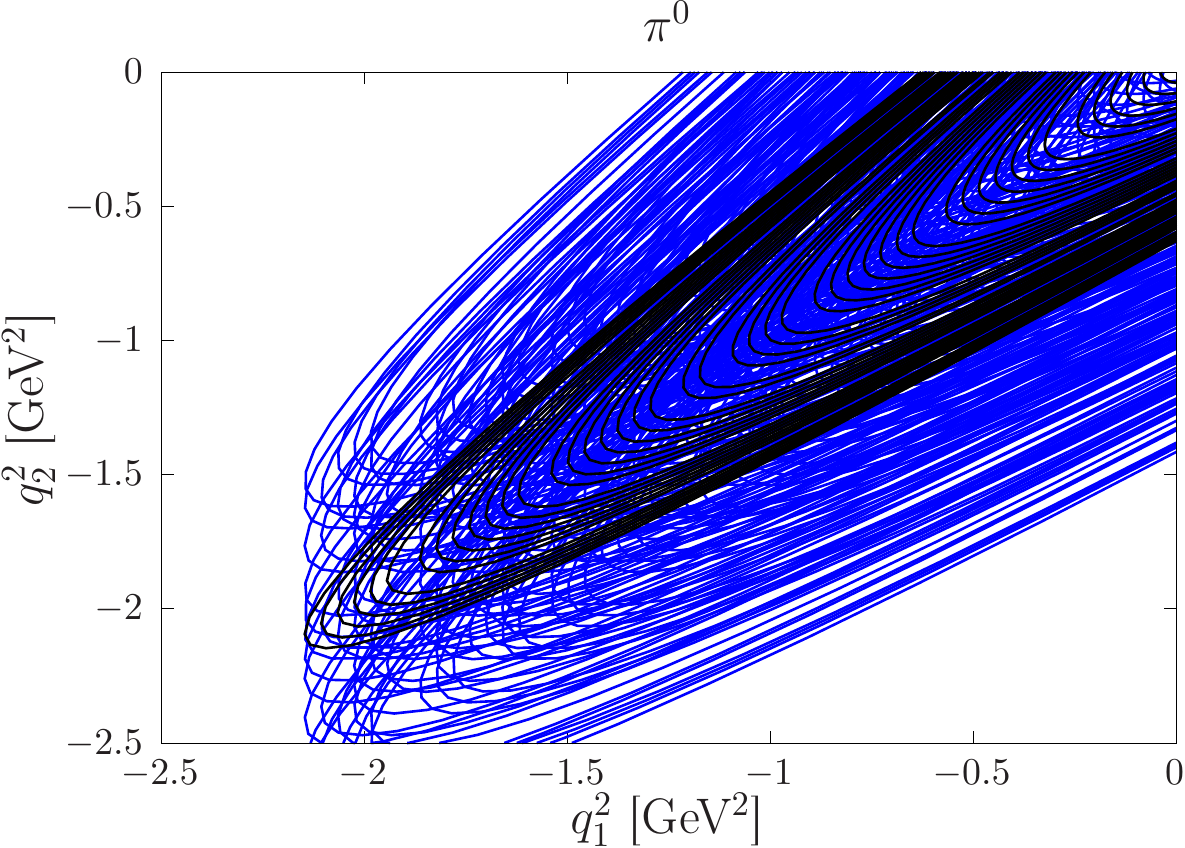}
	\includegraphics*[width=0.31\linewidth]{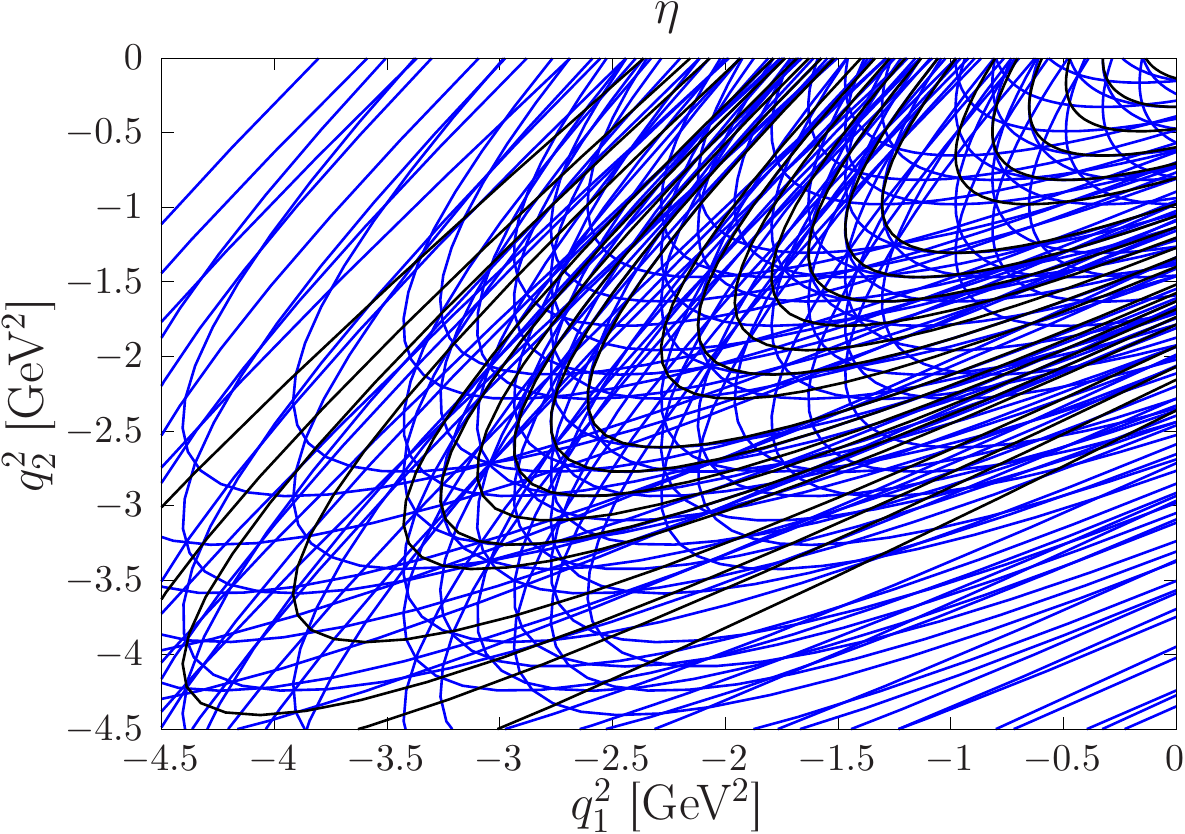}
	\includegraphics*[width=0.31\linewidth]{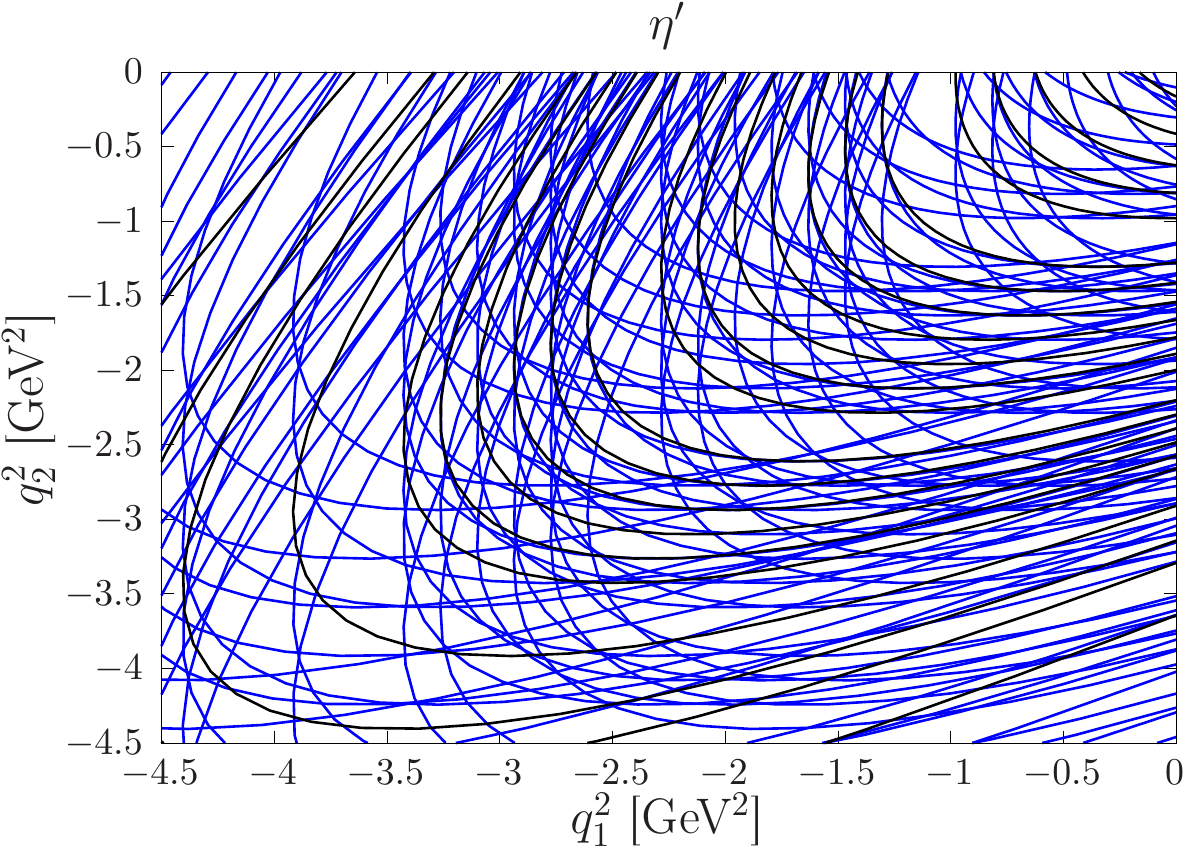}
	\vspace{-0.2cm}
	\caption{Kinematical reach for the $\pi^0$, $\eta$ and $\etap$ meson at the lattice spacing $a=0.0787~$fm. Each orbit corresponds to a given value of $|\vec{q}_1|$ and $|\vec{q}_2|$ and can be generated by continuously varying the parameter $\omega_1$. Black and blue lines correspond to the pseudoscalar rest frame and moving frame respectively.}	
	\label{fig:kin}
\end{figure}

Finally, all correlation functions have been computed for two values of the meson momentum ($\vec{p} = \vec{0}$ and $\vec{p} = (2\pi/L) \vec{z}$). The number of photon momenta ($\vec{q}_1$) is tuned on each ensemble to efficiently probe the ($Q_1^2, Q_2^2$) plane up to 2~GeV$^2$ for the pion and 4~GeV$^2$ for the $\eta$ and $\etap$ mesons, see \Fig{fig:kin}.

\section{Analysis \label{sec:analysis}}

In Ref.~\cite{Gerardin:2019vio}, the amplitude $\widetilde{A}_{\mu\nu}(\tau)$ was written in terms of a single scalar function $\widetilde{A}^{(1)}(\tau)$ and its derivative
\begin{equation}
\widetilde{A}_{\mu\nu}(\tau)  = -i Q^{E}_{\mu\nu} \ \widetilde{A}^{(1)}(\tau) + P^{E}_{\mu\nu} \ \frac{ \mathrm{d} \widetilde{A}^{(1)} }{\mathrm{d}\tau}(\tau) \,.
\label{eq:A1}
\end{equation}
The two coefficients $P^{E}_{\mu\nu} = i P_{\mu\nu}$ and $Q^{E}_{\mu\nu} = (-i)^{n_0} Q_{\mu\nu}$ do not depend on $\omega_1$ and are given by
\begin{equation}
q_{\mu\nu} \equiv \epsilon_{\mu\nu\alpha\beta} q_1^{\alpha} q_2^{\beta}  = P_{\mu\nu} \omega_1 + Q_{\mu\nu} \,.
\label{eq:qmunu}
\end{equation} 

In the analysis, all equivalent contributions to $\widetilde{A}^{(1)}$ are averaged over and we explicitly use the Bose symmetry relation
\begin{equation}
\label{eq:Bosym}
\widetilde{A}_{\mu\nu}(\tau; \vec{q}_1, \vec{q}_2) = \widetilde{A}_{\nu\mu}(-\tau; \vec{q}_2,\vec{q}_1) \, e^{-E_{P} \tau } \,.
\end{equation}
Once $\widetilde{A}^{(1)}(\tau)$, or equivalently $\widetilde{A}_{\mu\nu}(\tau)$ is known, the integration over $\tau$ in \Eq{eq:Mlat} is done using the Simpson rule. 
In \Section{sec:wick}, we will often display individual contributions to $\widetilde{A}^{(1)}(\tau)$ that originate from the different Wick contractions of~\Fig{fig:wick}. 
The scalar function $\widetilde{A}^{(1)}(\tau)$ is strongly affected by staggered oscillations due to the contribution of the parity partner state. We stress that these oscillations do not disappear as we take the continuum limit but the contribution of the parity partner state to the form factor does. Thus, for clarity, we will often display the smeared function
\begin{equation}
f(\tau)|_{\rm smr} = \frac{1}{4} f(\tau-a) + \frac{1}{2} f(\tau) + \frac{1}{4} f(\tau+a) \,,
\label{eq:smr}
\end{equation}
instead of $f$ itself. Since the smearing procedure and the numerical integration using the trapezoidal rule commute, integrating the smeared integrand does not impact the  sum over $\tau$.

\subsection{Wick contractions \label{sec:wick}}

\begin{figure}[t]
	\includegraphics*[width=0.49\linewidth]{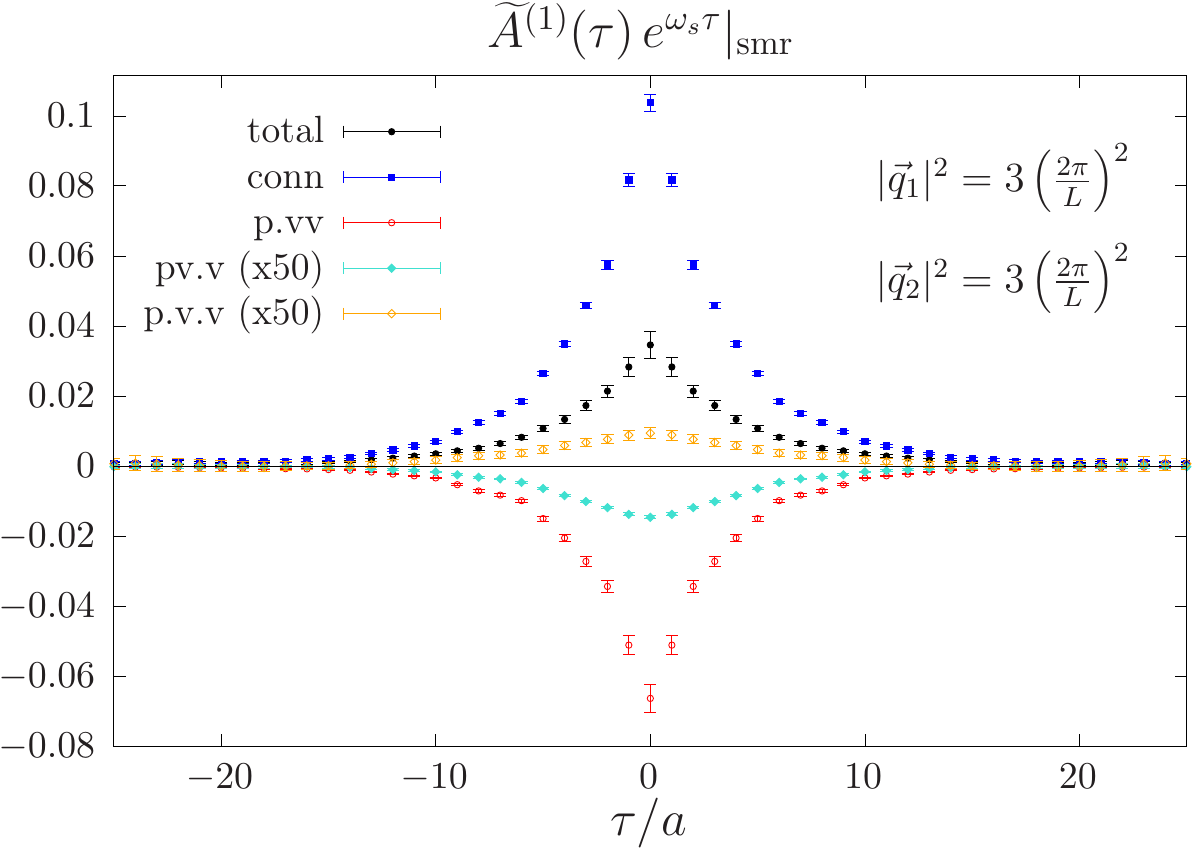}
	\includegraphics*[width=0.49\linewidth]{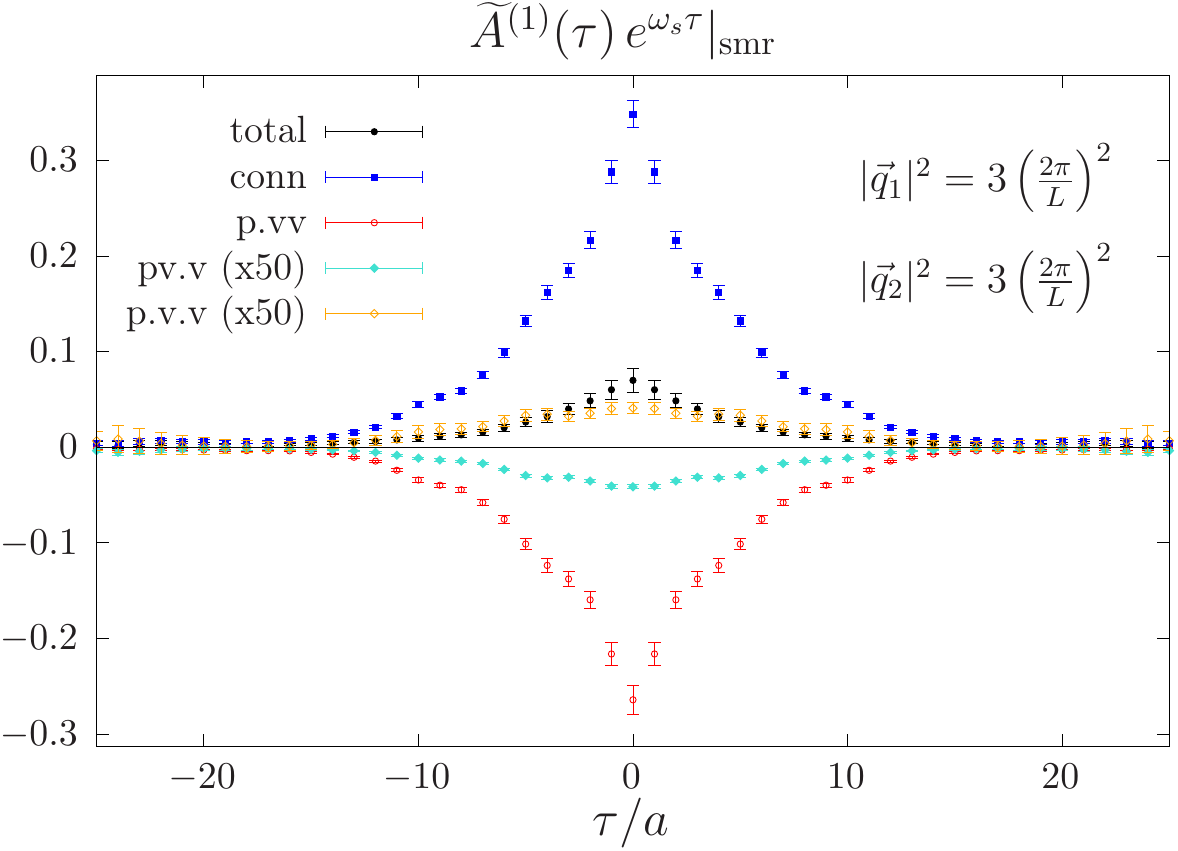}
	\caption{Contributions from the four Wick contractions, displayed in \Fig{fig:wick}, to the integrand in Eq.~(\ref{eq:Mlat}), for the $\eta$ (left) and $\etap$ (right) mesons. The figure corresponds to the choice $\omega_1 = \omega_s \equiv E_P/2$ that would lead to the double-virtual TFF with $Q_1^2=Q_2^2$. The smearing procedure is defined by \Eq{eq:smr}. The black circles represent the sum over all contributions. Some contributions have been rescaled by a factor 50 for clarity. The results are given at our finest lattice spacing.}	
	\label{fig:C}
\end{figure}

The contributions from the four Wick contractions of~\Fig{fig:wick} to the amplitude $\widetilde{A}^{(1)}(\tau)$ are shown in \Fig{fig:C} for one ensemble at our finest lattice spacing. 
In the isospin limit, the last two contractions, which involve a pseudoscalar loop, vanish for the pion and contribute only to the $\eta$ and $\etap$ transition form factors. 
In~\cite{Gerardin:2019vio,Feng:2012ck}, the authors have shown that the second diagram, with a single vector loop, is numerically subdominant. 
It is negative and contributes only at the percent level to the pion TFF. In particular, this diagram would exactly vanish in the SU(3) flavor limit.
For the same reason, the last diagram of~\Fig{fig:wick}, which contains two vector loops, is expected to give a very small contribution to the $\eta$ and $\etap$ form factors. Thus, for the $\eta$ and $\etap$ TFFs, most of the signal comes from the purely connected contribution and from the disconnected contribution that contains a single pseudoscalar loop. Both contributions have opposite sign and the cancellation is stronger for the $\etap$. 

\subsection{Tail correction} 

Because of the finite time extent of the lattice and of the increase of the statistical error, the integration over $|\tau|$, the time separation between the two vector currents, is truncated at $\tau_c \approx 1.5$~fm. 
In Ref.~\cite{Gerardin:2019vio}, the large-$\tau$ dependence of the amplitude was fitted assuming a vector meson dominance (VMD) or a lowest meson dominance (LMD) parametrization. This parametrization was then used to estimate the integrand above the cut. 
With staggered quarks, this procedure is more complicated due to the contribution of the parity partner state that appears with the usual factor $(-1)^\tau$. This contribution, responsible for the presence of oscillations at the level of the integrand, largely cancels between consecutive time slices and vanishes in the TFFs in the continuum limit. However, since we have little knowledge about the parity parter state contribution, and to avoid complicated fits, we decided to follow a different strategy. 
First, we compute the TFF neglecting the tail of the integrand above the cut. The resulting form factor is fitted using an LMD parametrization in the range $Q^2 \leq Q^2_{\rm tail}=1$~GeV$^2$ where the model provides an acceptable description of our data. 
In a second step, this parametrization is used to estimate the missing tail contribution at the level of the amplitude (we use the equations derived in Appendix A of Ref.~\cite{Gerardin:2019vio}). 
In principle, this method can be applied iteratively until it converges. 
In practice, since we use a conservative value of $\tau_c$, we find that at most two iterations already suffice at our level of precision. 
To avoid any bias due to this tail correction, we remove all points where the tail contributes more than 10\% to the integrand in~\Eq{eq:Mlat}.

\subsection{Finite-size effects}

\begin{figure}[t]
	\centering
	\includegraphics[width=0.4\textwidth]{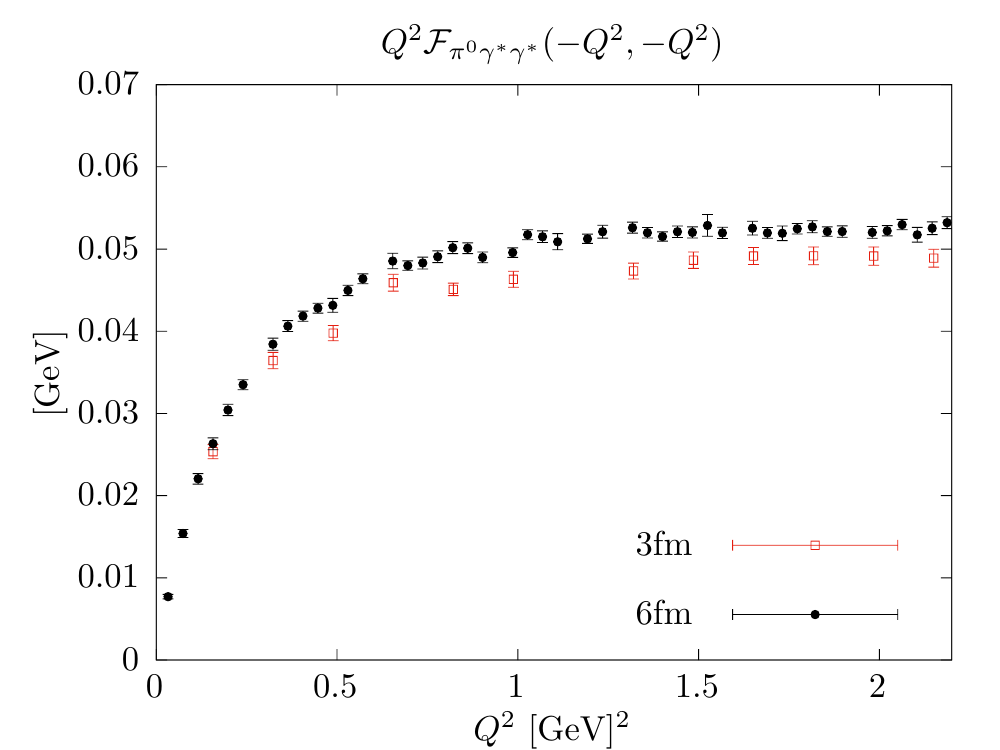}
	\includegraphics[width=0.4\textwidth]{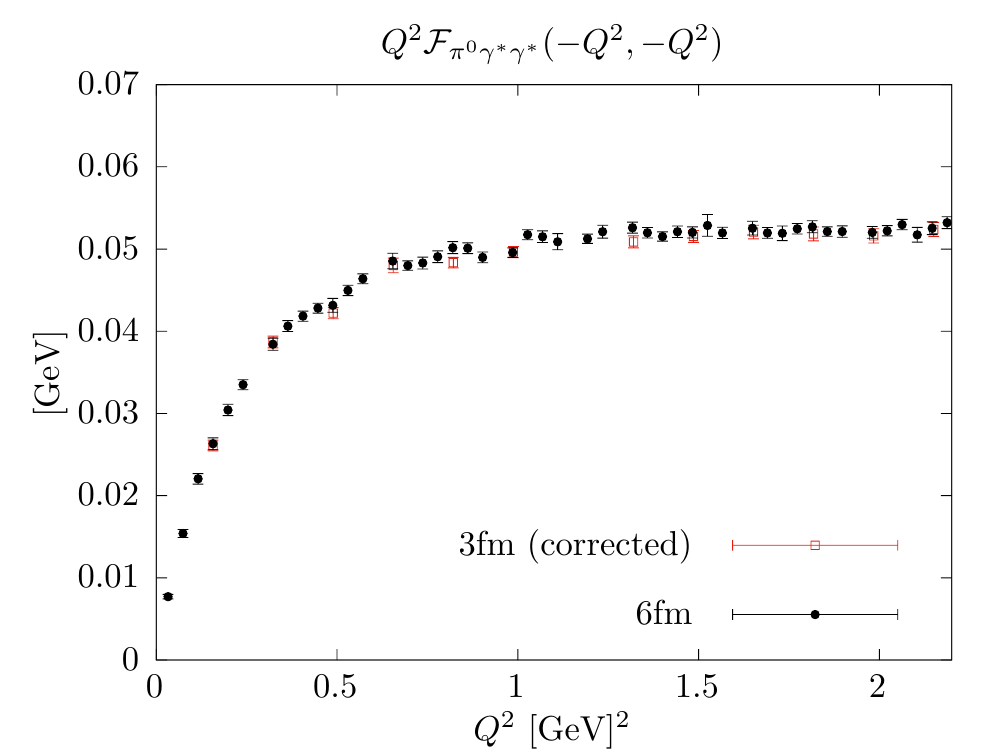}
	\caption{Pion TFF in the double-virtual regime using two different volumes ($L =3$~fm and $L=6$~fm), at our finest lattice spacing. Before (left) and after (right) correcting for backward propagating pion.}
	\label{fig:fse}
\end{figure}

The calculation of the disconnected diagram with a single pseudoscalar loop (third diagram in~\Fig{fig:wick}) is numerically expensive and dominates the cost of our simulations. 
In this section we study the possibility of using smaller physical volumes to increase the statistical precision. 

When using smaller volumes, two different kind of finite-size effects need to be considered. 
First, it was observed in Ref.~\cite{Gerardin:2016cqj} that the finite-time extent of the lattice potentially leads to large finite-time effects (FTE) due to backward propagating pions. Fortunately, a simple procedure was proposed to correct for this effect. Second, we need to consider finite-spatial-extent effects. 
This can be done by looking at the pion TFF for which a high statistical precision has been achieved on both large ($L=6$~fm) and small ($L=3$~fm) ensembles. 
Since the mass of the taste-singlet pseudoscalar meson decreases as we approach the continuum limit, we focus on our finest lattice spacing where the taste-singlet pion mass is approximately 200~MeV. The results for $L=6$~fm and $L=3$~fm boxes are depicted on \Fig{fig:fse} before and after FTE correction. In fact, when one corrects for FTE, the data for the two different box sizes agree within statistical errors, even at the lowest accessible virtualities $Q^2$. For the pion TFF, that is extracted exclusively from $L \approx 6~$fm ensembles, finite-size effects (FSE) are thus neglected.

Large volumes allow us to probe much smaller virtualities. 
This feature is very important for the pion-pole contribution, but to a lesser extent for the $\eta$- and $\etap$-pole contributions. First $\amueta$ and $\amuetap$ are less sensitive to the region close to the origin, see \Table{tab:contrib}. 
Second, as the pseudoscalar mass increases, lower virtualities can be reached, even in small volumes. 
To estimate the FSE due to the use of small spatial volumes at fine lattice spacings, the pion-pole contribution is computed from \Eq{eq:master} but using the weight functions of the $\eta^{(\prime)}$ meson. The relative difference between small and large volumes is then taken as our systematic.

\subsection{Excited-state contributions \label{sec:topt}} 

When using \Eq{eq:Amunu} to extract the $\eta$ form factor, instead of \Eq{eq:improved}, the $\etap$ is the first excited state and its relative contribution decays exponentially with $\Delta E \, t_P \equiv (E_{\etap} - E_{\eta}) t_P$
\begin{equation}
\widetilde{A}_{\mu\nu}^{(\eta); {\rm eff}}(\tau,t_P)  =\widetilde{A}_{\mu\nu}^{(\eta)}(\tau) \left( 1 + \frac{E_{\eta}}{E_{\etap}} \frac{ Z_8^{(\etap)} }{ Z_8^{(\eta)} }  \frac{ \widetilde{A}_{\mu\nu}^{(\etap)}(\tau) }{ \widetilde{A}_{\mu\nu}^{(\eta)}(\tau) }  e^{-\Delta E \, t_P}  + \cdots \right) \,,
\end{equation}
where the ellipsis represents higher excited state contributions. 
The ratio of overlap factors strongly depends on the choice of interpolating operator used for the $\eta$ and, at the physical point, the mass gap is $\Delta E \approx 400$~MeV. 
With staggered quarks, because of taste breaking effects, the mass gap between the two taste-singlet mesons decreases at large lattice spacings: we find $\Delta E \approx 300$~MeV at our coarsest lattice spacing. 
With our choice of interpolating operator, and at the physical $\eta^{(')}$ masses, we find that $t_P \geq 1.0~$fm  would be needed to reduce the exponentially suppressed  term to the level of 10\% and 1.35~fm for 5\% (the target precision if one aims at 10\% uncertainty on the pseudoscalar-pole contribution to $\ahlbl$). 
We conclude that using \Eq{eq:Amunu} instead of \Eq{eq:improved} would require large values of $t_P$

\begin{figure}
	\centering
	\includegraphics[width=0.455\textwidth]{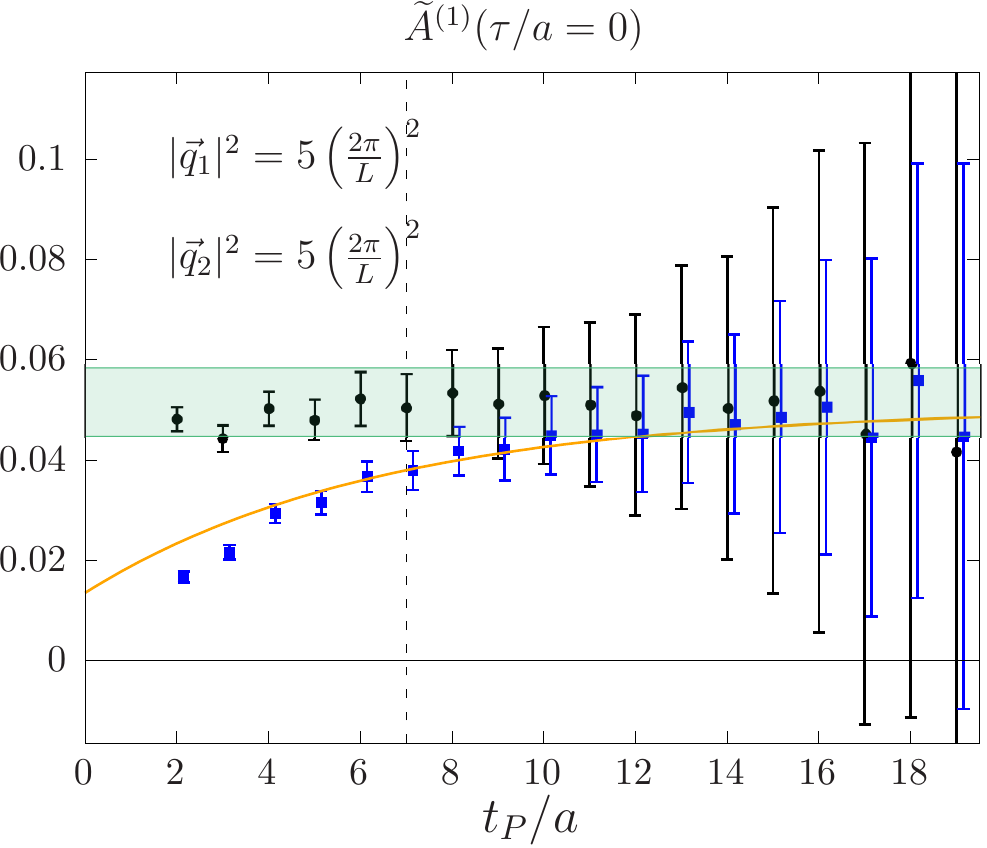} \quad
	\includegraphics[width=0.46\textwidth]{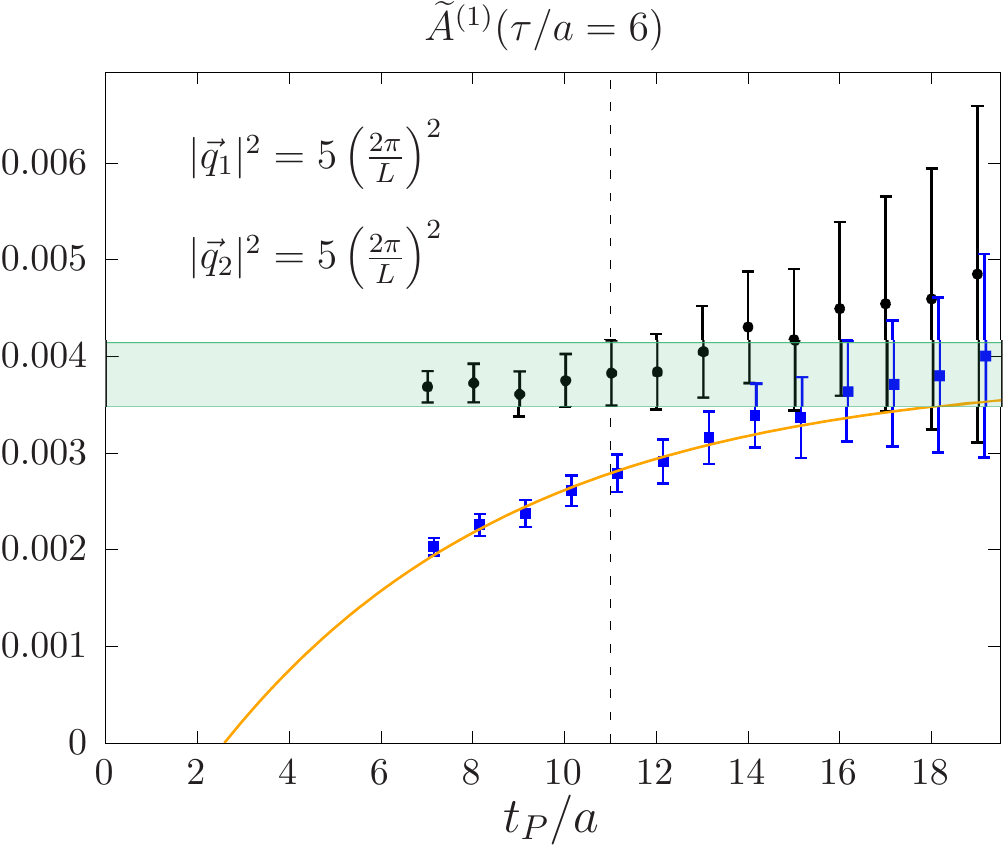}
	
	\caption{Effective value for the function $\widetilde{A}^{(1)}(\tau)$ as a function of $t_P$ for the $\eta$ meson at the lattice spacing $a=0.0787~$fm. The horizontal green band corresponds to our optimal choice. Black circles are obtained using the improved estimator given by \Eq{eq:improved}, while the blue squares are the corresponding data using the estimator in \Eq{eq:Amunu}. The orange line is the excited state prediction for the estimator (\ref{eq:Amunu}) as predicted by \Eq{eq:dec}. Blue points are slightly shifted to the right for clarity. }
	\label{fig:topt1}
\end{figure}

However, despite the use of AMA and large statistics, the disconnected diagram that contains a single pseudoscalar loop is noisy and becomes the main source of statistical error. Thus, it is very useful to use as small values of $t_P$ as possible. If, instead of  \Eq{eq:Amunu}, one extracts the TFFs from a matrix of correlation functions, the $\etap$ contribution to the $\eta$ TFF is essentially suppressed (up to statistical precision in the determination of the overlap factors). 
In \Fig{fig:topt1}, we compare the effective value of $\widetilde{A}^{(1)}(\tau = 0)$ as a function of $t_P$ using both methods. The second approach, in black, where the $\etap$ contribution is effectively removed, leads to a plateau at much earlier times. 
The value of $t_P$, for each value of $\tau$, is chosen in the plateau region to minimize the statistical uncertainty while keeping systematic errors under control (vertical line in \Fig{fig:topt1}). For ensembles at the same $\beta$, the same values of $t_P$ are employed while for different values of $\beta$, the values of $t_P$ are approximately the same in physical units.
In practice, $\widetilde{A}^{(1)}(\tau)$ is obtained from a smearing over three values of $t_P$ using the same weights factors as in \Eq{eq:smr}.

\begin{figure}
	\centering
	\includegraphics[width=0.40\textwidth]{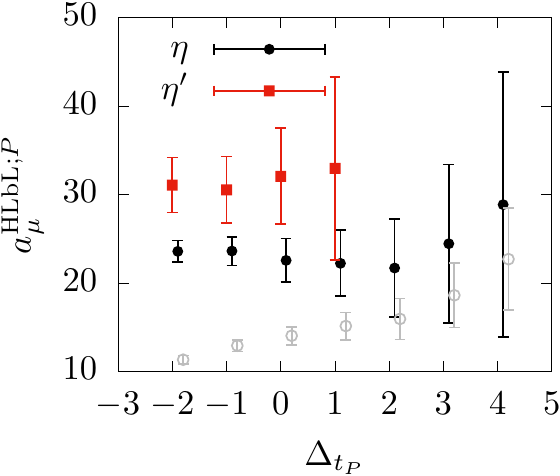}
	\includegraphics[width=0.40\textwidth]{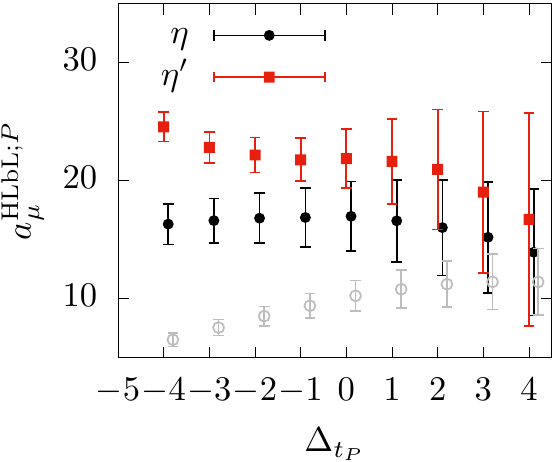}
	\caption{Value of $\amu$ as a function of $\Delta t_P$ using the improved estimator. The point $\Delta t_P=0$ corresponds to the choice used in our analysis. Left: $a=0.1315$~fm. Right: $a=0.0787$~fm. The open gray circles are the results for the $\amueta$ using the estimator (\ref{eq:Amunu}).}
	\label{fig:topt2}
\end{figure}

In \Fig{fig:topt2}, we present two stability plots where the pseudoscalar-pole contribution for the $\eta$ (black) and $\etap$ (red) are shown as a function of $\Delta t_P$, a constant shift from our optimal values for $t_P$ (negative values mean more aggressive choices of $t_P$). With the improved estimator, we hardly see any trend, except for the $\etap$ at fine lattice spacing. 
For comparison, we also present the result for the $\eta$ meson using the naive estimator (in gray).
In this case, we observe a clear slope due to excited-state contamination, and both methods to extract $\amueta$ eventually agree at large $t_P$ although the statistical error increases significantly.

\subsection{Treatment of statistical and systematic errors \label{sec:syst}}

Statistical errors are estimated using the jackknife method after blocking. All observables are computed on the same set of gauge field configurations and using the same block size such that we can easily propagate statistical errors. The lattice spacing was computed in~\cite{Borsanyi:2020mff}, with a relative precision at the permil level, but using a different blocking procedure. The lattice spacing enters as a trivial factor in the conversion of the TFF and the virtualities in physical units but also in the continuum extrapolations. We have generated pseudosamples with a Gaussian distribution to propagate the error on the lattice spacing.  
This error is always subdominant so that it is safe to neglect the correlation of the lattice spacing with the quantities studied here.

During the analysis several choices are made. First, to extract the form factors on individual ensembles, we need to choose $\Delta t_P$ (for the $\eta$ and $\etap$ mesons, see \Section{sec:topt}), the fitting range $Q^2_{\rm tail}$ for the LMD or VMD model used for the tail correction, the values of $\omega_1$ that are used to samples the orbits (see \Eq{eq:Mlat} and above) and the strategy to compute the pseudoscalar energies and overlap factors (see \Section{sec:systerror}). 
Second, in the continuum extrapolation, we can choose the order of the $z$-expansion described in \Section{sec:zexp}, add higher order terms in the continuum extrapolation or perform cuts in lattice spacings. 
Several analyses have been performed by varying all these choices. Since some fits do not include all correlations, the systematic error is estimated assuming a flat weight among these representative variations. An explicit example is given in \Section{sec:systerror} where we compute the pseudoscalar-pole contributions.

\section{Results \label{sec:results} }

\subsection{Transition form factors at a single lattice spacing} 

\begin{figure}
	\centering
	\includegraphics[width=0.38\textwidth]{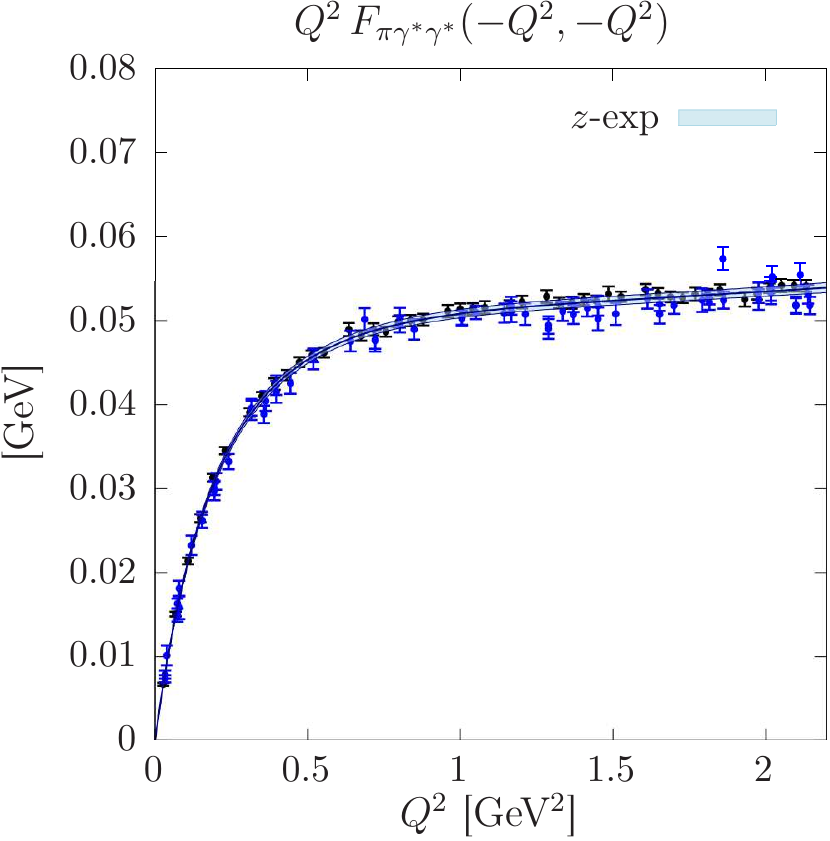} \quad
	\includegraphics[width=0.38\textwidth]{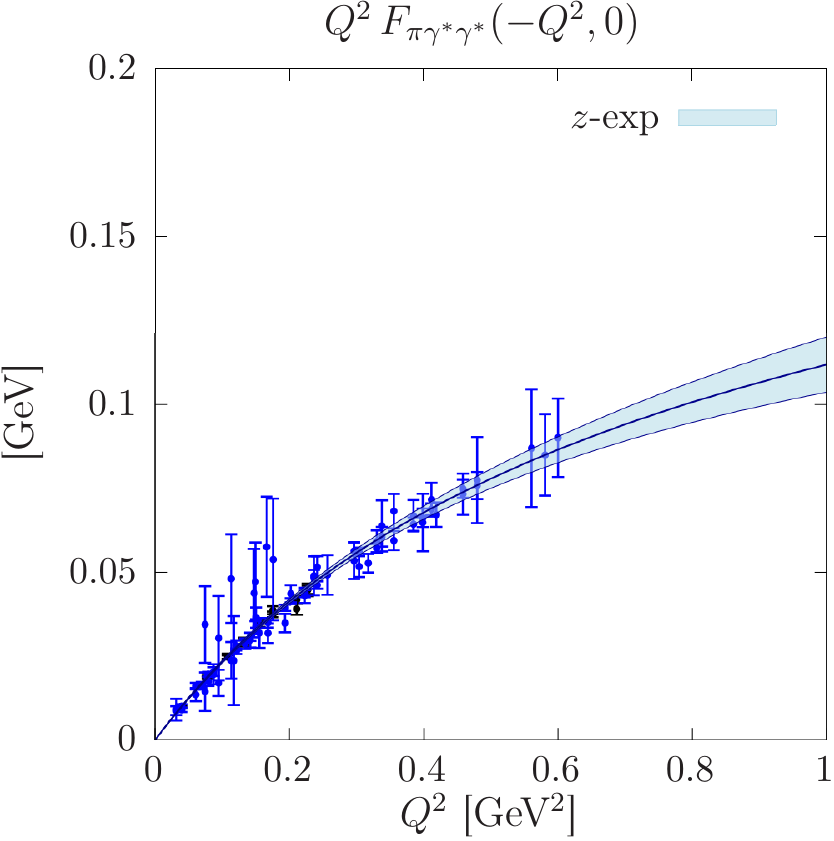}
	\vspace{0.3cm}

	\includegraphics[width=0.38\textwidth]{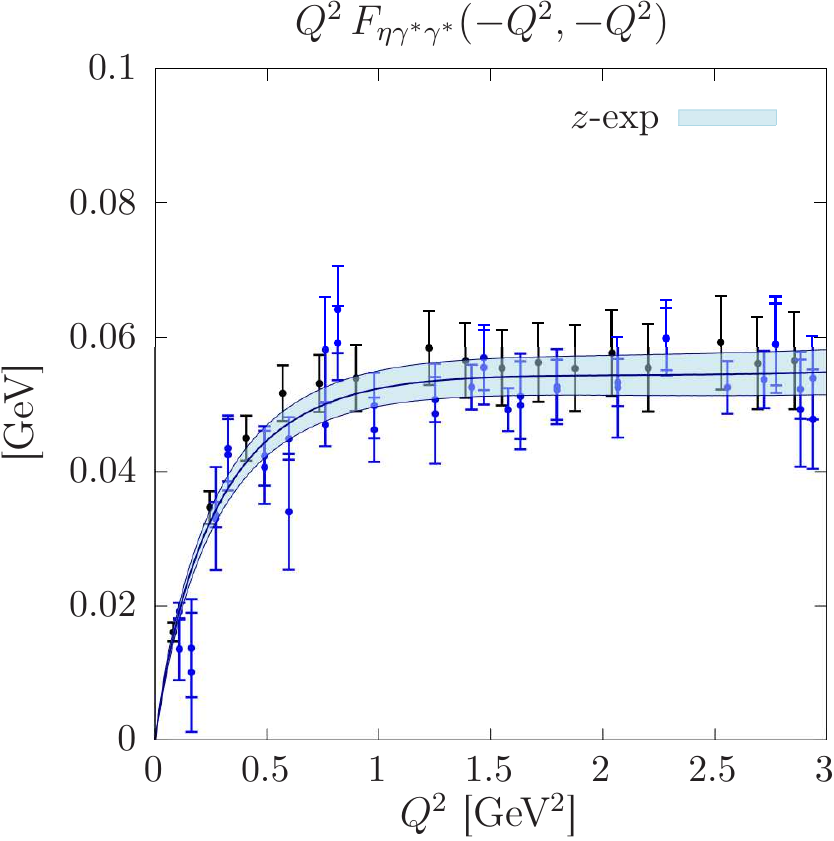} \quad
	\includegraphics[width=0.38\textwidth]{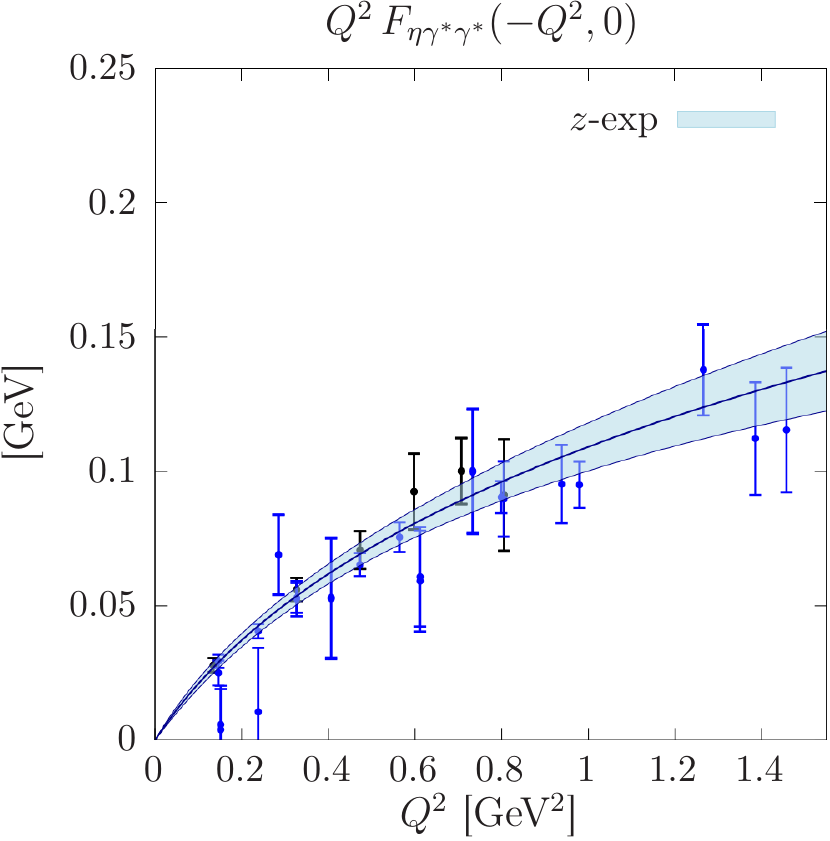}		
	\vspace{0.3cm}

	\includegraphics[width=0.38\textwidth]{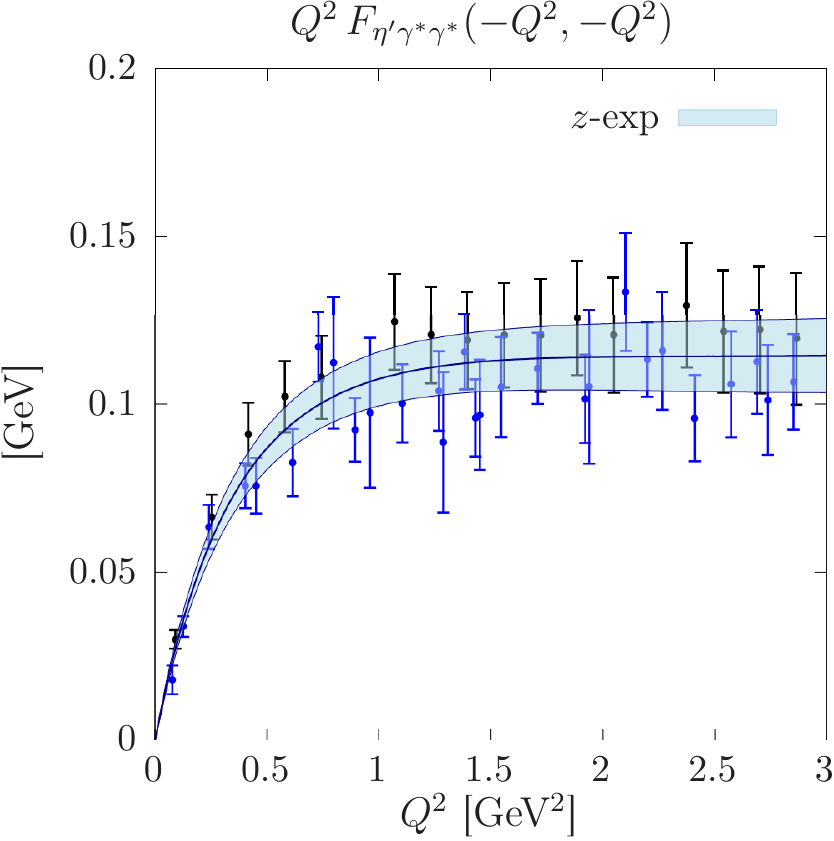} \quad
	\includegraphics[width=0.38\textwidth]{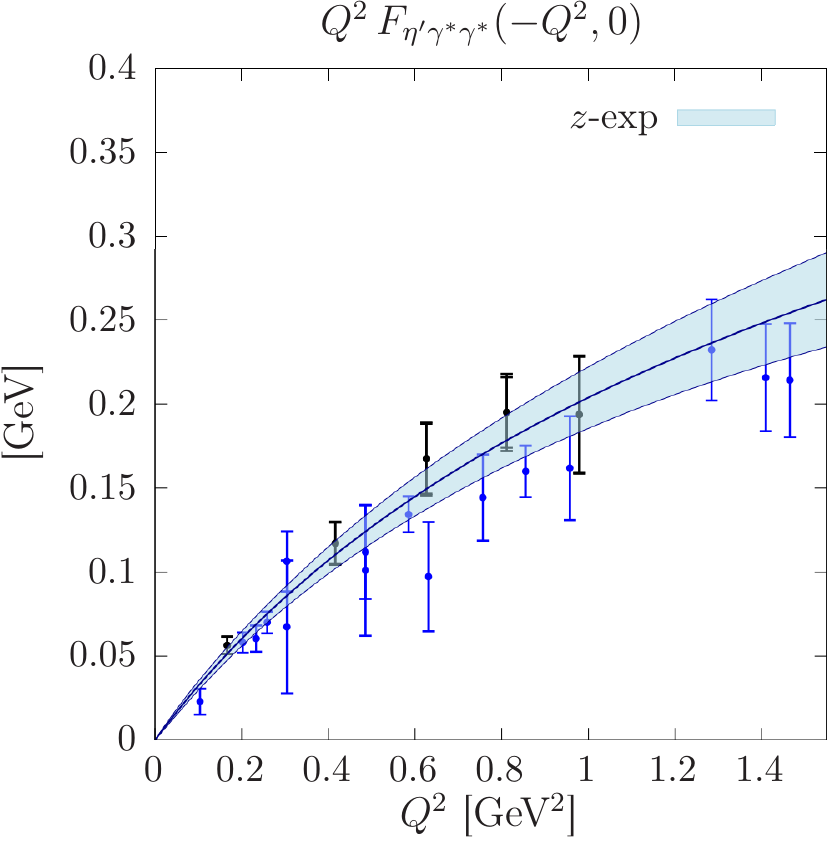}			
		
	\caption{TFF of the $\pi^0$ (top panel), $\eta$ (second panel) and $\etap$ (bottom panel) in the double (left) and single (right) virtual regimes. Black filled circles and blue filled circles indicate, respectively, $\vec{p}=\vec{0}$ and $\vec{p} = \frac{2\pi}{L}(0,0,1)$. The blue bands are fit based on the parametrization given by \Eq{eq:zexp} at a single lattice spacing ($a=0.0787~$fm). }
	\label{fig:singlea}
\end{figure}

Results for the TFFs, at a single lattice spacing ($a=0.0787$~fm), are presented in \Fig{fig:singlea}. For clarity, we present our results for two specific kinematics: the single-virtual form factor with one real photon and the double-virtual form factor where both photons share the same virtuality $Q_1^2=Q_2^2$. In both cases, we observe a good agreement between the pseudoscalar rest frame and moving frame. In~\cite{Gerardin:2019vio} the motivation for adding a  nonvanishing pseudoscalar momentum was to better probe the ($Q_1^2,Q_2^2$) plane. This is crucial for the pion because of its low mass (left panel in \Fig{fig:kin}). For the heavier $\eta$ and $\etap$, a good coverage of the ($Q_1^2,Q_2^2$) plane, even for the single-virtual TFF, is already reached in the meson rest frame.

For the pion, the signal essentially comes from the fully connected contribution with a small negative contribution, at the percent level, from the second disconnected diagram in \Fig{fig:wick} (see Appendix~\ref{sec:discV}). Thus, a very high statistical precision can be achieved. Nevertheless, the signal-to-noise ratio deteriorates at low virtualities. This is of particular relevance when computing the pseudoscalar-pole contribution to $a^{\rm HLbL}_{\mu}$ for which the low-Q$^2$ region dominates. This point is further discussed in \Section{sec:g-2}. 
The signal for the $\eta$ and $\etap$ TFFs is noisier for two reasons. First, the disconnected diagram with a single pseudoscalar loop (third diagram in \Fig{fig:wick}) is large and enters with an opposite sign as compared to the fully connected contribution. Second, the overlap factors and the energies of the pseudoscalar mesons also contribute to the error.

In the following section, we extrapolate our data to the continuum limit.

\subsection{Parametrization of $\FF$ and extrapolation to the physical point \label{sec:zexp}} 

To extrapolate the form factor to the continuum limit, we fit our lattice data using the modified $z$-expansion introduced in~\cite{Gerardin:2019vio,Boyd:1995sq}
\begin{multline}
P(Q_1^2,Q_2^2) \ \FF(-Q_1^2, -Q_2^2)  = \\  \sum_{n,m=0}^{N} c_{nm}(a) \, \left( z_1^n + (-1)^{N+n} \frac{n}{N+1} \, z_1^{N+1} \right) \, \left( z_2^m +  (-1)^{N+m} \frac{m}{N+1} \, z_2^{N+1} \right) \,,
\label{eq:zexp}
\end{multline}
where $z_k$ are conformal variables
\begin{align}
z_k = \frac{\sqrt{t_c + Q_k^2} - \sqrt{t_c - t_0 \vphantom{Q_k^2}}}{\sqrt{t_c + Q_k^2} + \sqrt{t_c - t_0\vphantom{Q_k^2}}}, \quad k = 1,2,
\end{align}
$c_{nm}$ are coefficients symmetric in $m$ and $n$, $t_c = 4m_\pi^2$ maps the branch cut of the TFF onto the unit circle $\left| z_k \right|= 1$ and $t_0$ is a free parameter. 
The choice $t_0 = t_c ( 1 - \sqrt{1+ Q_{\max}^2 / t_c } )$ 
ensures that the maximum value of $\left|z_k\right|$ is minimized in the momentum range $Q^2<Q_{\max}^2$. 
 Finally, the function $P(Q_1^2,Q_2^2)$ is an arbitrary analytic function. In~\cite{Gerardin:2019vio} it was argued that a convenient choice is 
\begin{equation}
P(Q_1^2,Q_2^2) = 1 + \frac{Q_1^2 + Q_2^2}{\Lambda^2} \,,
\end{equation}
such that the form factor parametrization decreases as $1/Q^2$ at large virtualities, even at finite value of $N$, as expected from the operator product expansion (OPE)~\cite{Lepage:1979zb,Lepage:1980fj,Brodsky:1981rp,Nesterenko:1982dn,Novikov:1983jt}. For the pion and the $\eta$ TFFs, $\Lambda=775~$MeV is set to the $\rho$ meson mass. For the $\etap$, we use $\Lambda=1~$GeV. In practice, we do not observe a strong dependence on this parameter. 

With staggered quarks, we expect discretization errors to enter quadratically~\cite{Lepage:1997id,Lepage:1998vj}. They are taken into account by expanding the $c_{nm}$ coefficients as
\begin{equation}
c_{nm}(a) = c_{nm}(0) \left( 1 + \gamma_{nm} (a \Lambda_{\rm QCD})^2 \right)\,.
\end{equation}
with $\Lambda_{\rm QCD} = 0.5~\GeV$ taken as a typical QCD scale. 
The data for the pion is the most precise. In this case, to check for possible higher order discretization effects and to estimate the associated systematic error, additional fits that include a quartic term $\delta_{nm} a^4$ have been performed. However, to avoid overfitting, those additional parameters are included only when the error on the corresponding quadratic coefficient, $\gamma_{nm}$, is well below 50\%. For the pion, the data are also limited to virtualities $Q^2 \lesssim 2.2~\GeV^2$. To stabilize the fit at large virtualities, the additional constraint $Q^2 \FFpi(-Q^2, -Q^2) = 0.062(15)~$GeV at $Q^2=20~\GeV^2$ is imposed. We believe that this 25\% uncertainty is conservative as it is larger than the difference between the TFF value at $Q_1^2=Q_2^2=2~\GeV^2$ and the asymptotic value~\cite{Lepage:1979zb,Lepage:1980fj,Brodsky:1981rp}
\begin{equation}
Q^2 \FFpi(-Q^2, -Q^2)  \xrightarrow[Q^2 \to \infty]{} \frac{2 F_{\pi}}{3}
\end{equation} 
predicted by the OPE. For the pion decay constant in the chiral limit, we use $F_{\pi} = 92.4~$MeV.

The main analysis is done using $N=2,3$ for the pion and for the $\eta$ meson, and with $N=1,2$ for the less precise $\etap$ TFF.
As can be seen in \Fig{fig:singlea}, the number of points included in the fits is very large. 
Furthermore, many ensembles are fitted simultaneously. It is clear that a fully correlated fit is not feasible. Thus, we decided to perform uncorrelated fits but to estimate the error through the jackknife procedure. This strategy prevents us from giving a clear meaning to the fit quality, but leads to reliable error estimates~\cite{Ammer:2019yjp,Bruno:2022mfy}. 

\begin{figure}[t!]
	\includegraphics[width=0.49\linewidth]{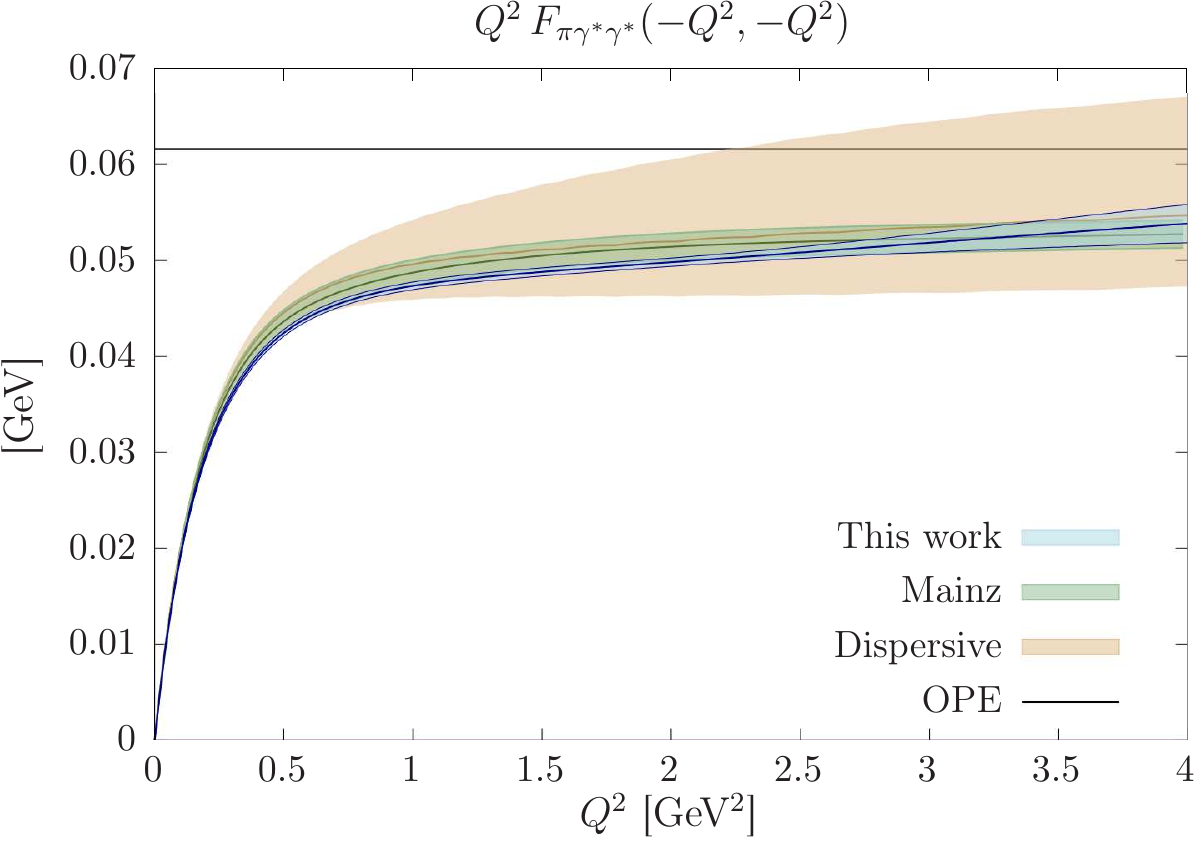}
	\includegraphics[width=0.49\linewidth]{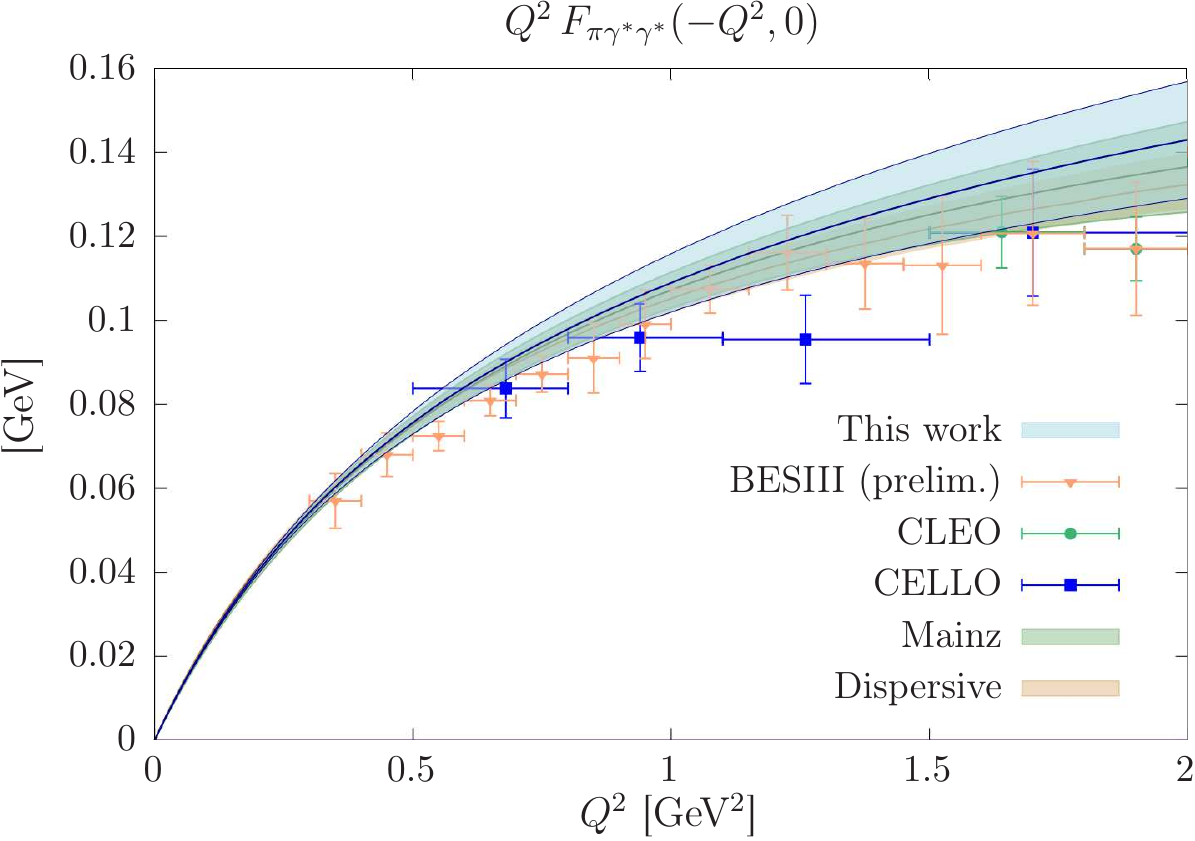} 	
	\vspace{0.3cm}
	
	\includegraphics[width=0.49\linewidth]{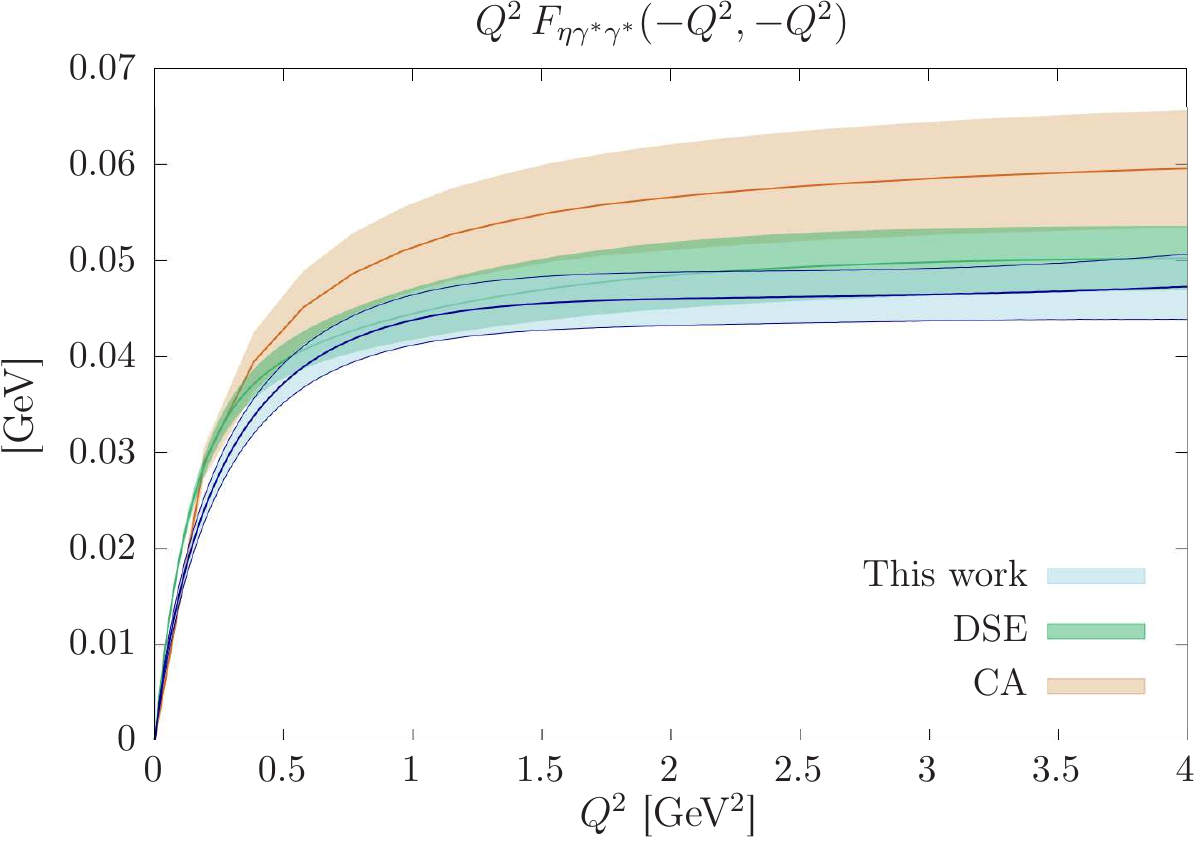}
	\includegraphics[width=0.49\linewidth]{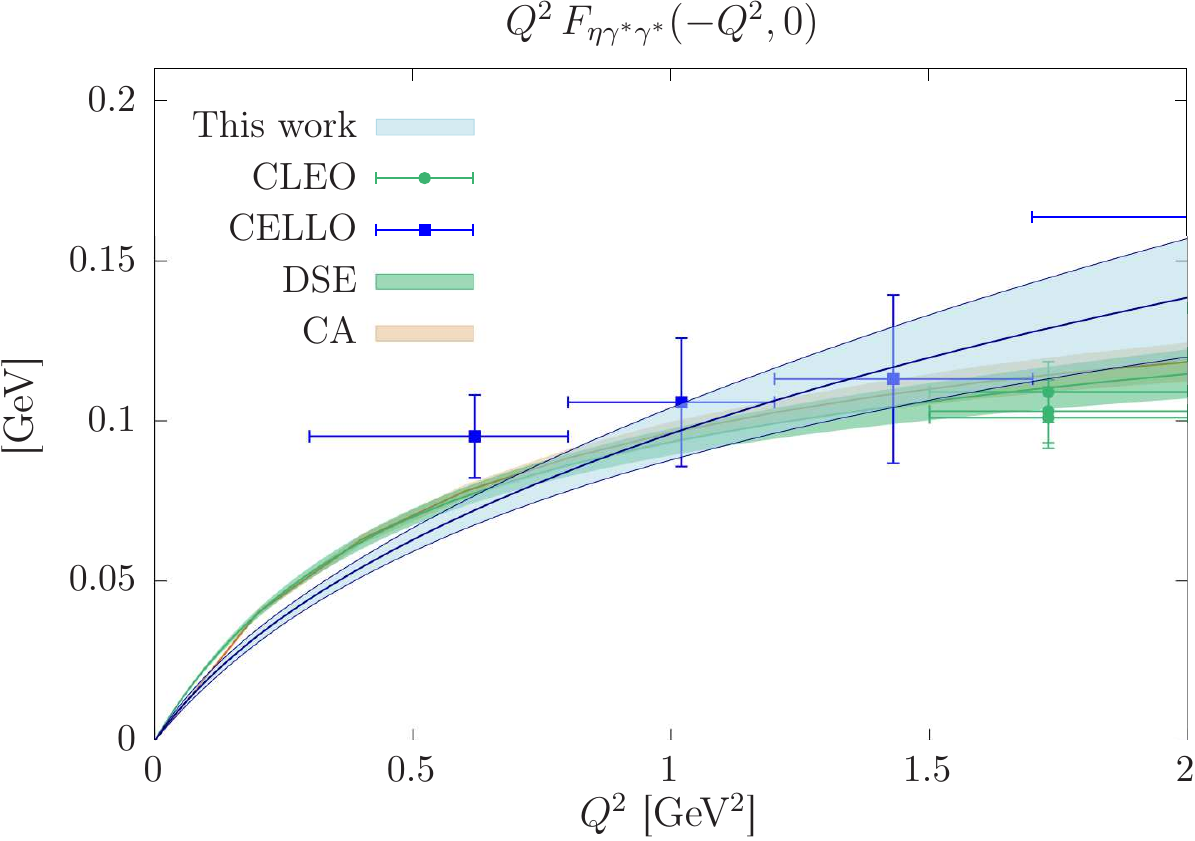}	
	\vspace{0.3cm}
	
	\includegraphics[width=0.49\linewidth]{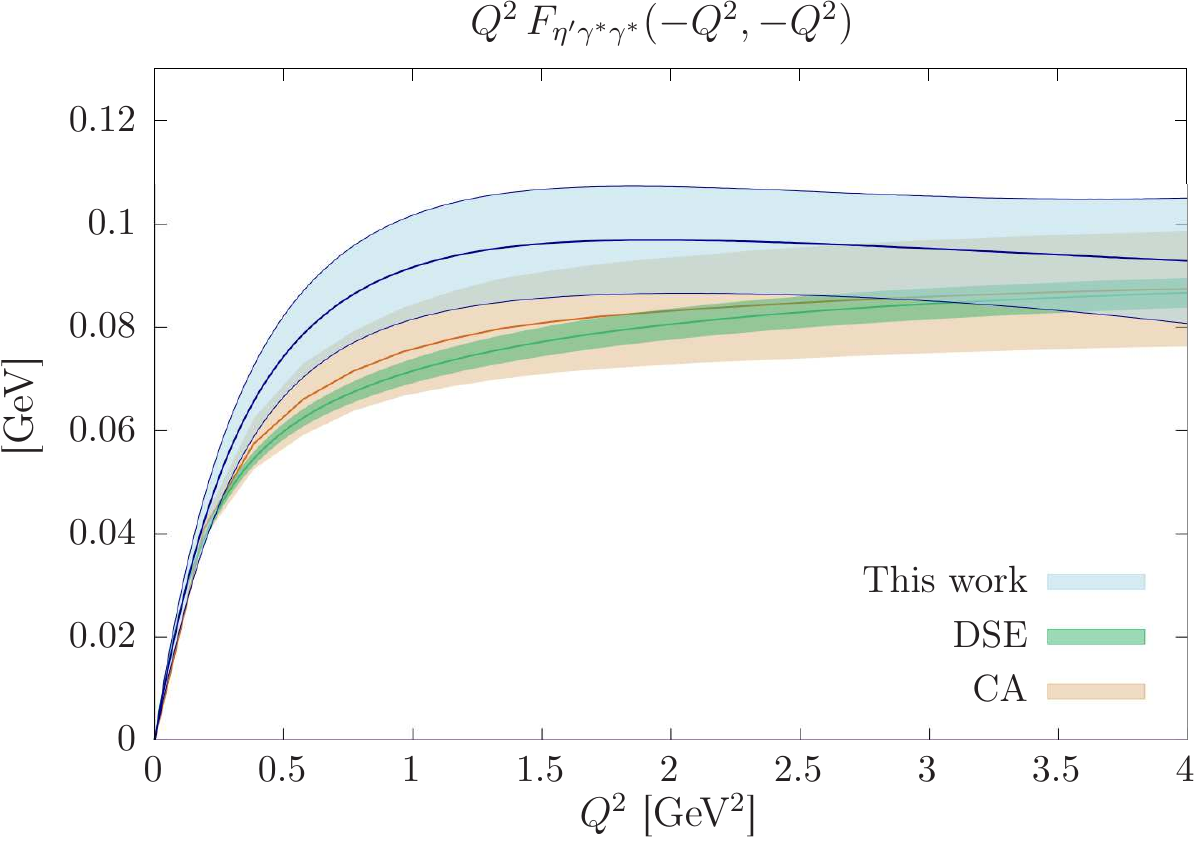}
	\includegraphics[width=0.49\linewidth]{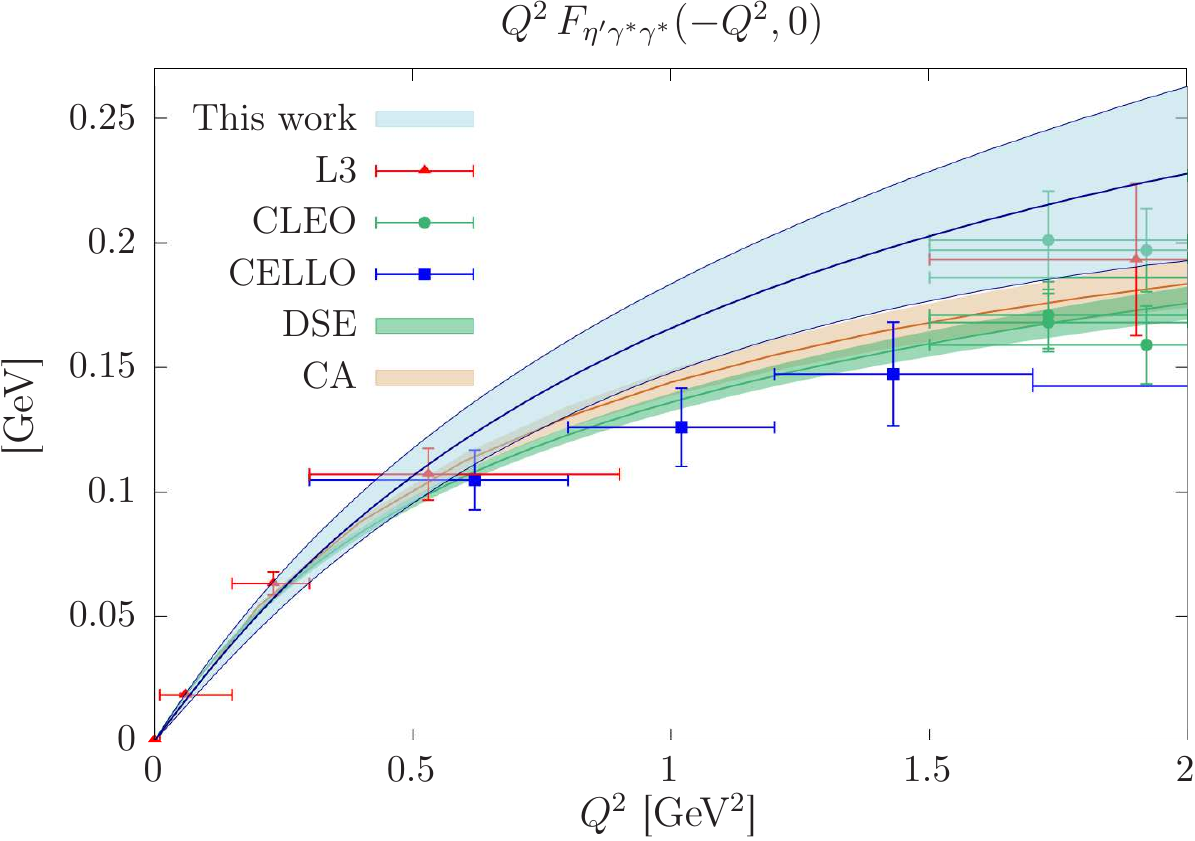}	
	
	\caption{TFFs at the physical point using the $z-$expansion from \Eq{eq:zexp}. Errors are statistical only. The horizontal black line corresponds to the OPE prediction. The dispersive result is extracted from~\cite{Hoferichter:2018dmo}, the Canterbury appoximant (CA) result from~\cite{masjuan:2017tvw} and the Dyson-Schwinger equation (DSE) result comes from~\cite{Eichmann:2019tjk}. Measurements from CELLO~\cite{CELLO:1990klc}, CLEO~\cite{CLEO:1997fho} and L3~\cite{L3:1997ocz} are shown for comparison in the single-virtual case. The preliminary results from the BES III experiment have been extracted from~\cite{Aoyama:2020ynm}}
	\label{fig:phys}
\end{figure}

\begin{figure}[t!]
	\includegraphics*[width=0.32\linewidth]{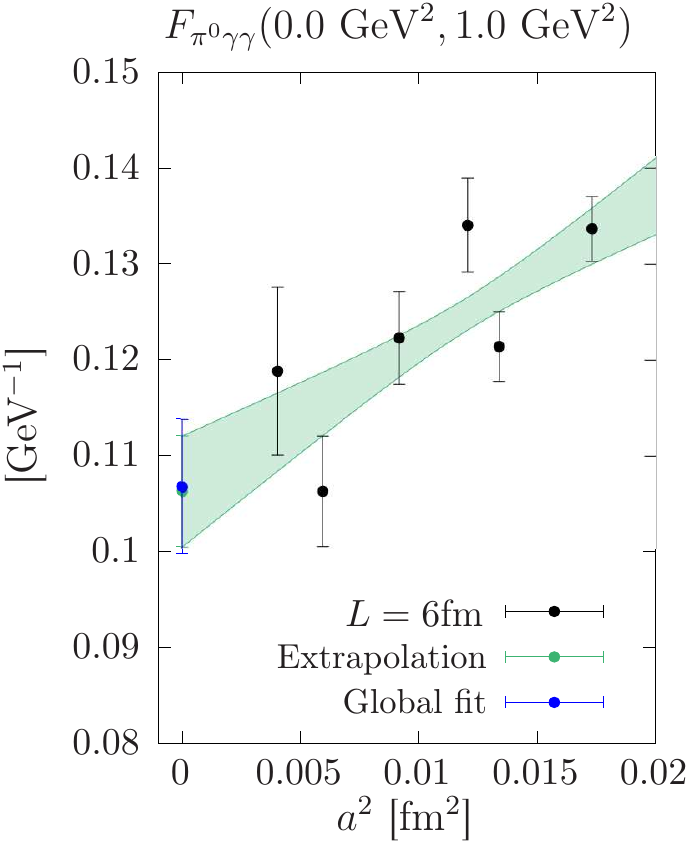}
	\includegraphics*[width=0.32\linewidth]{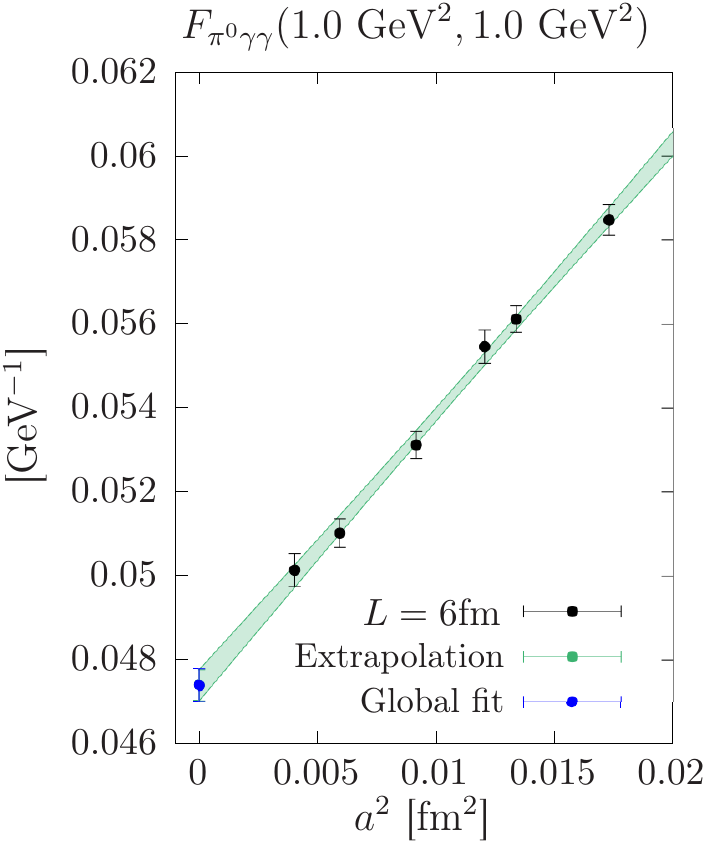}
	\includegraphics*[width=0.32\linewidth]{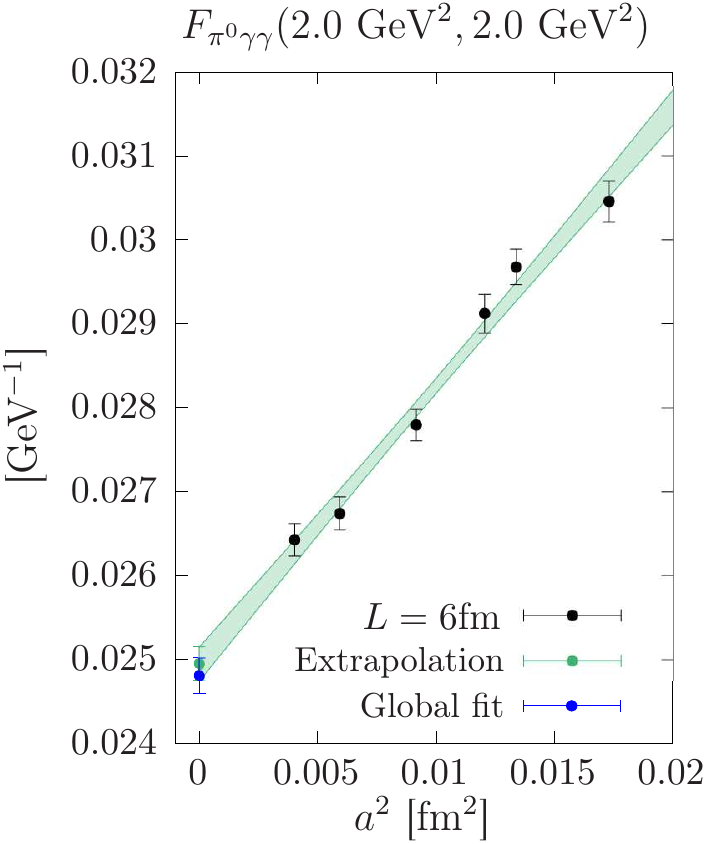}	
	\vspace{0.3cm}
	
	\includegraphics*[width=0.32\linewidth]{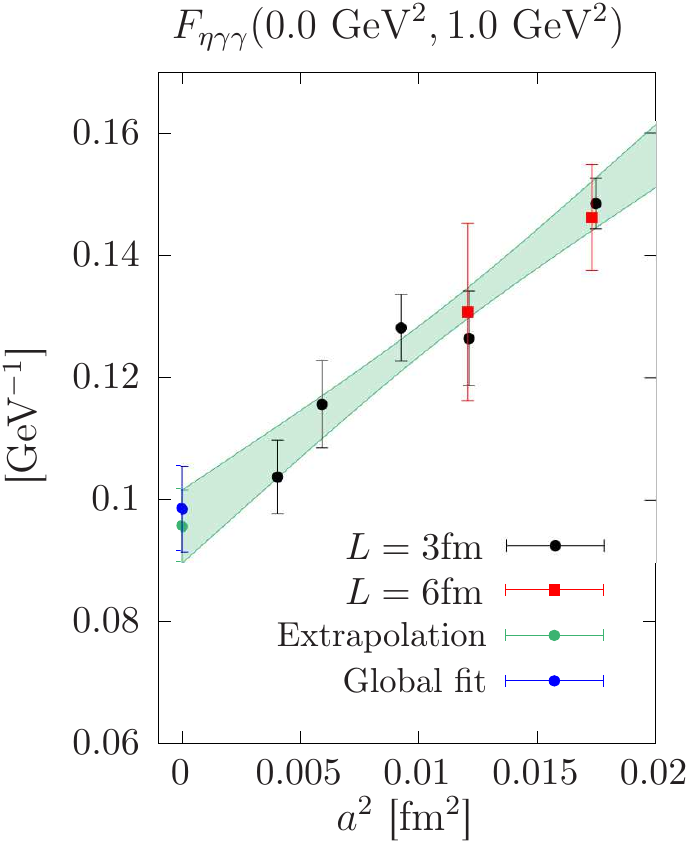}
	\includegraphics*[width=0.32\linewidth]{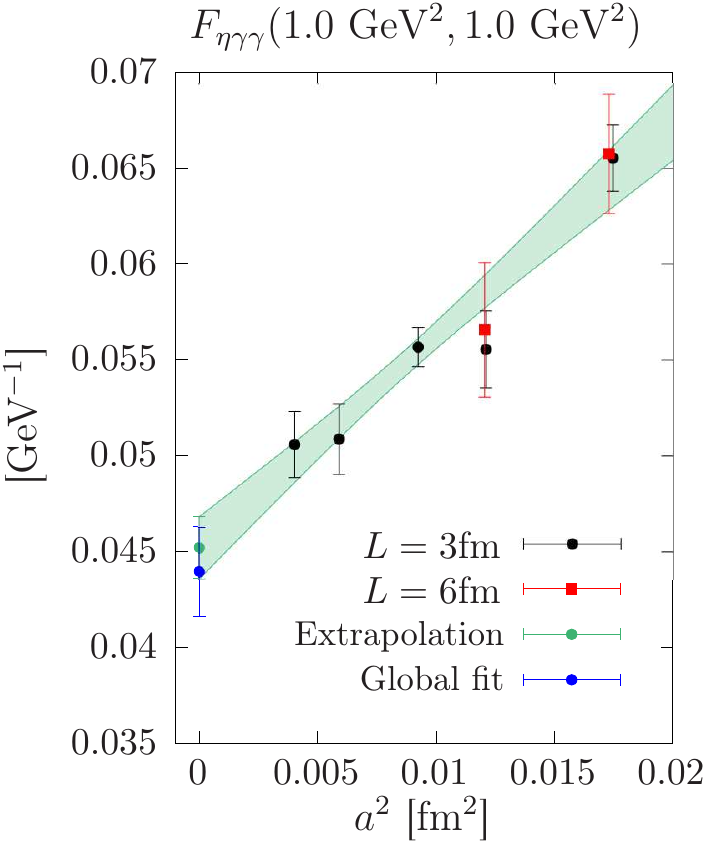}
	\includegraphics*[width=0.32\linewidth]{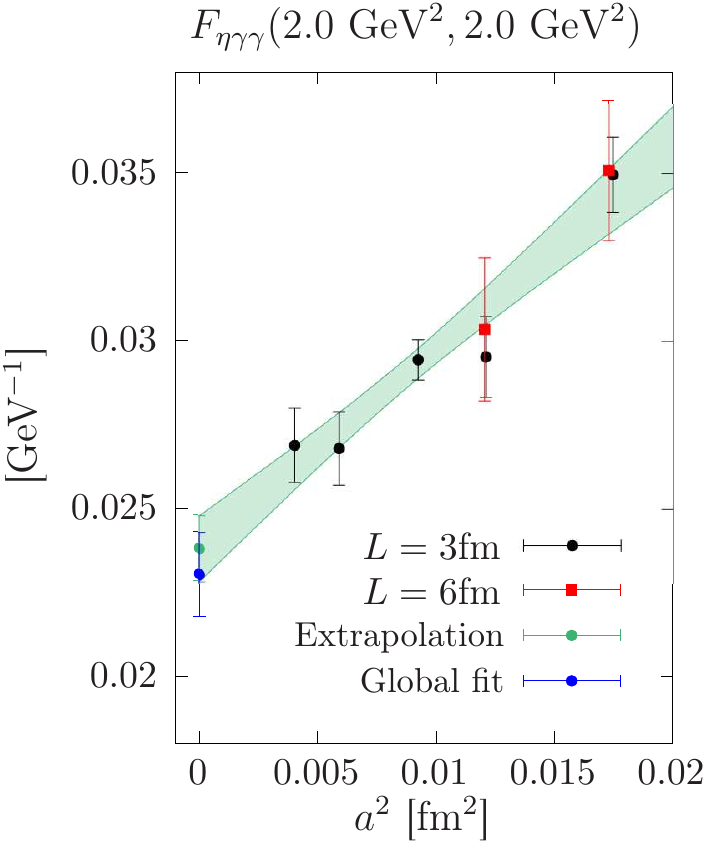}	
	\vspace{0.3cm}
	
	\includegraphics*[width=0.32\linewidth]{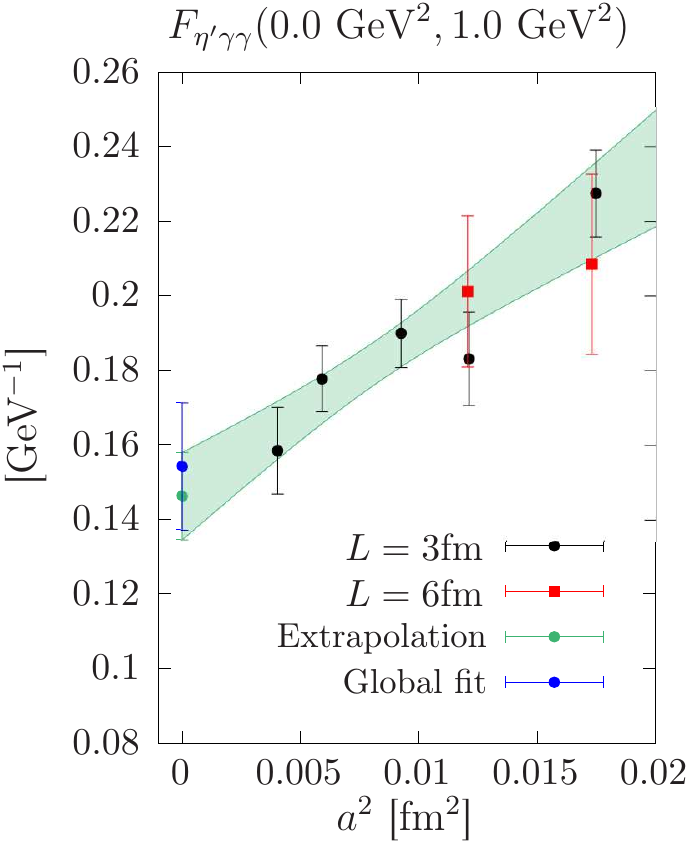}
	\includegraphics*[width=0.32\linewidth]{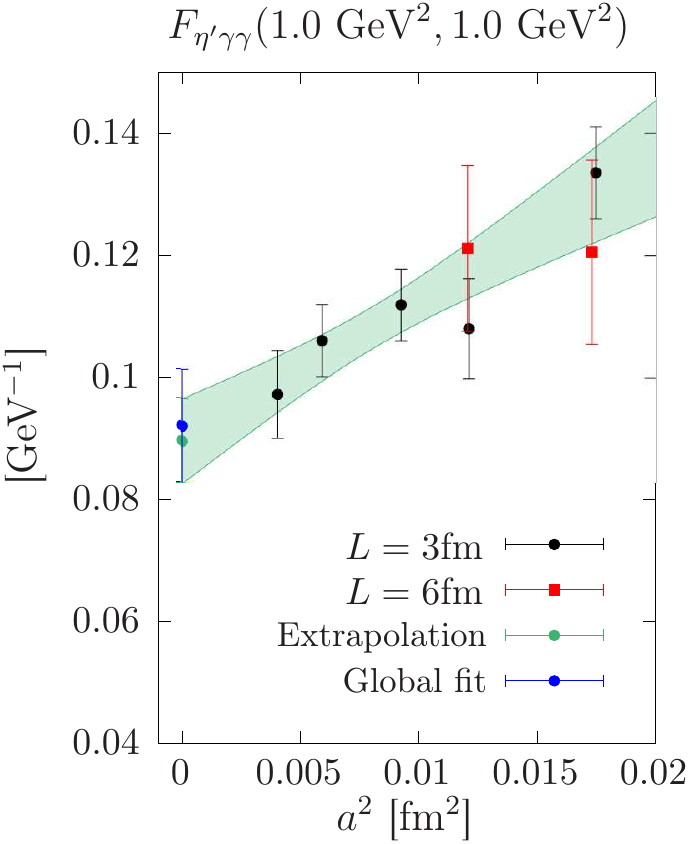}
	\includegraphics*[width=0.32\linewidth]{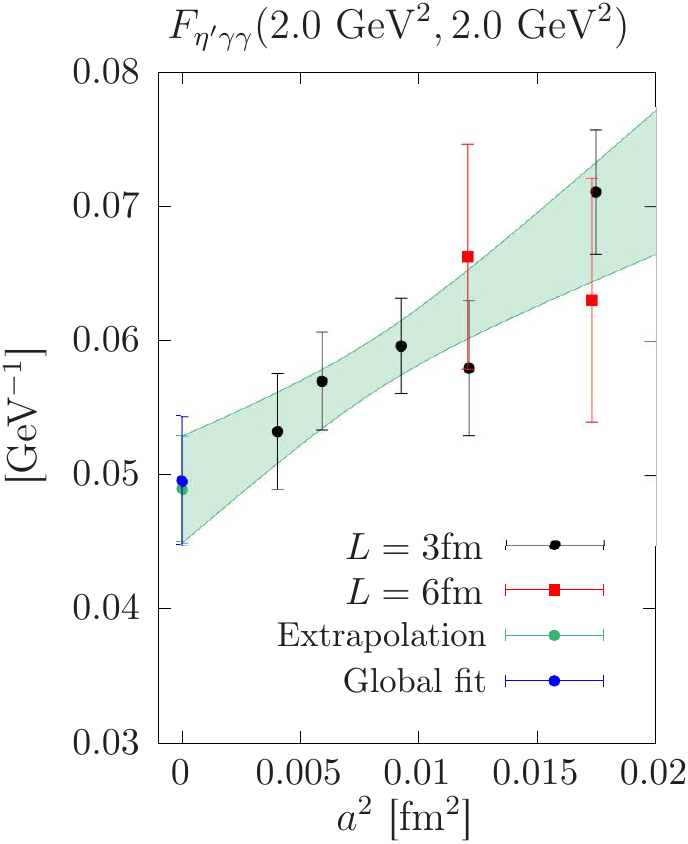}
	
	\caption{Continuum extrapolations of the TFFs for three kinematics $(Q_1^2,Q_2^2)$. Black and red points are obtained by fitting individual ensembles using~\Eq{eq:zexp}. The blue points are the result of the global fit procedure described in the text. Errors are statistical only.}
	\label{fig:TFFextrap}
\end{figure}

The final results are depicted in \Fig{fig:phys}. The fit parameters corresponding to one of our systematic variations (see \Section{sec:syst}), and the associated correlation matrix, for each pseudoscalar, are given in Appendix~\ref{app:tables}. 
The uncorrelated chi-squared per degree of freedom of the global fits are typically 1.15, 0.95 and 0.85 for the $\pi^0$, $\eta$ and $\etap$ cases. We note that $\chi^2/\dof \approx 1$ are still expected as the global fits include uncorrelated ensembles.  In~\Fig{fig:TFFextrap}, we show the continuum extrapolation of the TFFs for three different sets of virtualities $(Q_1^2,Q_2^2)$. In this case, each ensemble is fitted individually using~\Eq{eq:zexp} and the extrapolation is compared with the global fit procedure. In all cases, we observe a quadratic scaling in the lattice spacing. The normalization of the TFFs are discussed in the following section. 

In the single-virtual case, our result for the $\eta$ TFF shows a slight tension with the smallest CELLO bin. This feature is reflected by the tension in the normalization for the TFF discussed in the next section. For the $\etap$ meson, our result is slightly above the CELLO data although the tension is well below two combined standard deviations. The pion TFF is in very good agreement with experimental measurements. 
For the pion and $\etap$ TFFs we find reasonable agreement with previous determinations based on lattice calculations of the pion TFF~\cite{Gerardin:2019vio}, the dispersive framework~\cite{Hoferichter:2018dmo}, Canterbury approximants~\cite{masjuan:2017tvw} and Dyson-Schwinger equations~\cite{Eichmann:2019tjk}. In the case of the $\eta$ TFF, we observe a tension below $0.4~\GeV^2$ that is explained by our small value of two-photon decay rate as compared to experiments.

\subsection{Normalization of the form factors \label{sec:norm}}

When both photons are real, the form factor is related to the decay rate of the pseudoscalar meson to two photons. At leading order in QED
\begin{equation} 
\Gamma(P \to \gamma\gamma) = \frac{\pi \alphaQED^2 m_{P}^3}{4} \mathcal{F}^2_{P\gamma\gamma}(0, 0) 
\end{equation}
where $\alphaQED$ is the fine-structure constant. The neutral pion decay width has been measured by the PrimEx~\cite{PrimEx:2010fvg} and PrimEx-II~\cite{PrimEx-II:2020jwd} experiments with a precision of 1.5\% leading to the combined result $\Gamma(\pi^0 \to \gamma\gamma) = 7.802(52)_{\rm stat}(105)_{\rm syst}$~eV. It translates to a relative precision of 0.75\% on the normalization of the form factor. 
From the PDG average~\cite{Workman:2022ynf}, the $\eta$ and $\etap$ decay widths are $\Gamma(\eta \to \gamma\gamma) = 516 \pm 18$~eV and $\Gamma(\etap \to \gamma\gamma) = 4.28 \pm 0.19$~keV. It corresponds to a relative precision of 1.7\% and 2.2\%, respectively. 
In the case of the $\eta$ meson, a significant tension exists between the Primakoff measurement~\cite{Browman:1974sj} and the results from collider experiments~\cite{JADE:1985biu,CrystalBall:1988xvy,Roe:1989qy,Baru:1990pc,KLOE-2:2012lws}. A new measurement by PrimEx-Eta is expected in the comings years~\cite{primexeta}. 

\begin{figure}[t]
	\includegraphics*[width=0.32\linewidth]{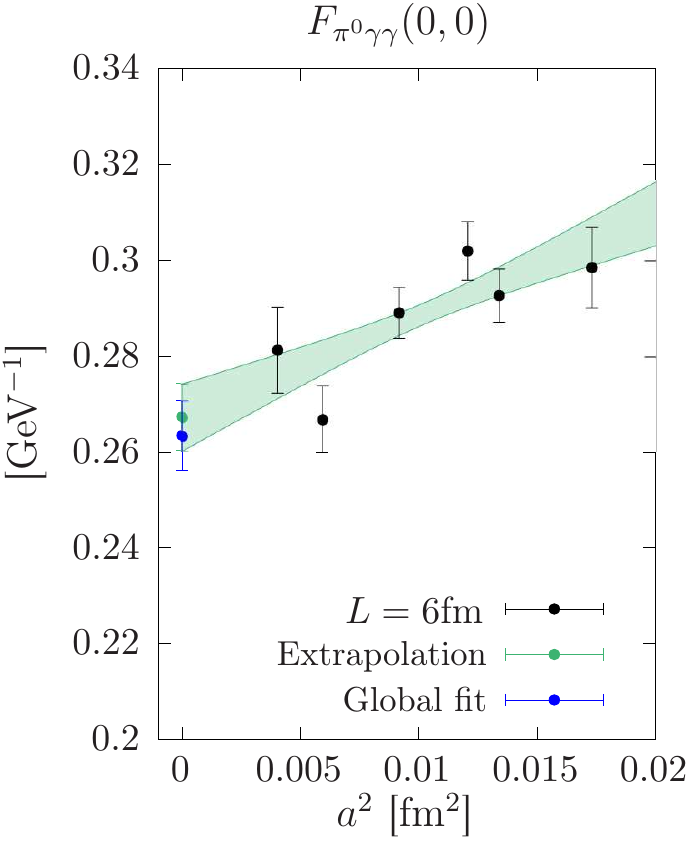}
	\includegraphics*[width=0.32\linewidth]{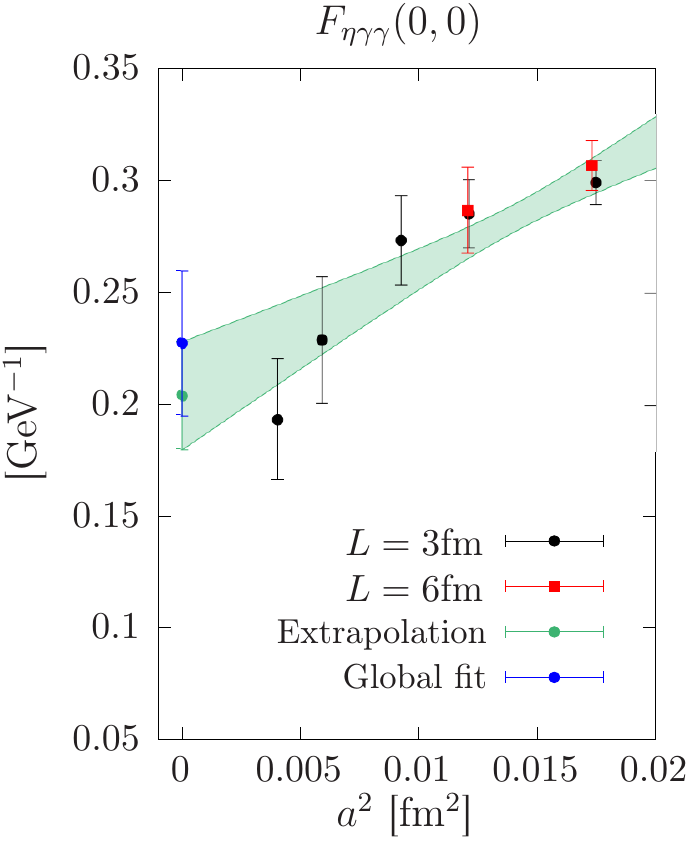}
	\includegraphics*[width=0.32\linewidth]{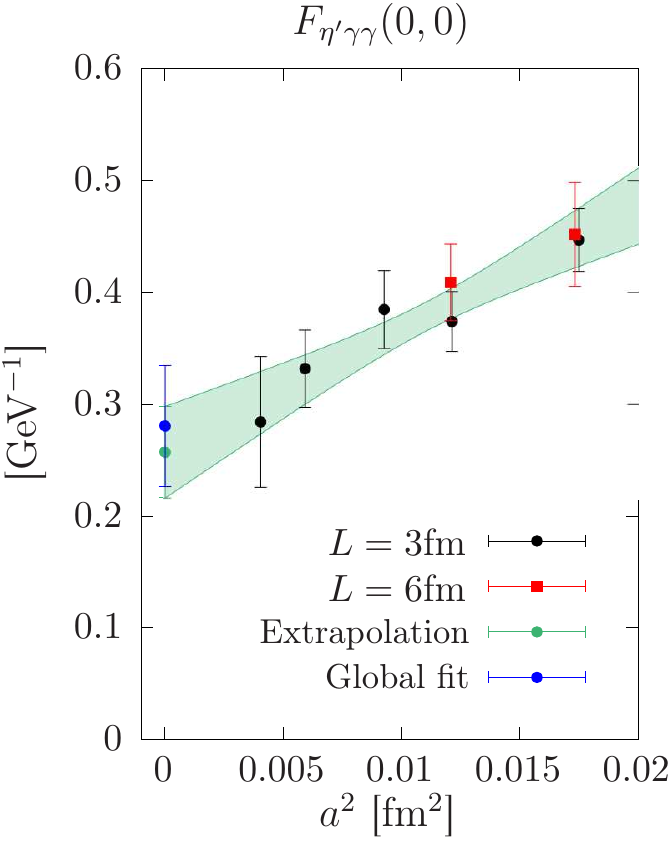}
	\caption{Continuum extrapolation of the normalization of the form factor $\FF(0,0)$ from one typical variation described in \Section{sec:syst}. The blue points are the results obtained from the global fit strategy. Black and red points are obtained by fitting the form factor on each ensemble separately. The corresponding extrapolation is then shown in green. For the $\eta$ and $\etap$ mesons, the red points correspond to large-volume ensembles only.}
	\label{fig:alpha}
\end{figure}

From our global fit procedure, described in \Section{sec:zexp}, we obtain
\begin{subequations}
\begin{align}
\Gamma(\pi^0 \to \gamma\gamma) &= 7.11 \, \pm \, 0.44_{\rm stat} \, \pm \, 0.21_{\rm syst}~\eV \\
\Gamma(\eta^{\ } \to \gamma\gamma) &= 338 \, \pm \, 94_{\rm stat} \, \pm \, 35_{\rm syst}~\eV \\
\Gamma(\etap \to \gamma\gamma) &= 3.4 \, \pm \, 1.0_{\rm stat} \, \pm \, 0.4_{\rm syst}~\keV  \,.
\end{align}
\end{subequations}
Our results for the pion and the $\etap$ mesons are compatible with the experimental values at the level of 1.4 and 0.8 standard deviation respectively. For the $\eta$ meson, our value presents a slight tension at the level of 1.7 standard deviations and is smaller in magnitude. The impact on the pseudoscalar-pole contribution will be discussed in the following section. 

We note that, at our finer lattice spacings, our results are based on relatively small physical volumes. Furthermore, FSE are expected to be the largest at small virtualities and might affect the normalization of the form factor more strongly. 
In addition to the global fit, we have also performed individual fits on each ensemble separately. The large volume ensembles, shown in red in \Fig{fig:alpha}, fully support the small decay rate for the $\eta$ meson. Furthermore, we remind the reader that we do not see any significant FSE for the pion where the calculation has been performed on both large and small volumes at all lattice spacings. 
We also tried to fit our data using the LMD model described in~\cite{Gerardin:2019vio} and we obtain a fully compatible result, $\Gamma(\eta^{\ } \to \gamma\gamma) = 301 \, \pm \, 50_{\rm stat}$~eV. 
Finally, we note that the ETM result $\Gamma(\eta^{\ } \to \gamma\gamma) = 323 \pm 88~\eV$~\cite{Alexandrou:2022qyf} is also smaller than the experimental value based on collider experiments. However, this result is based on a single ensemble and it is difficult to predict the size of the discretization effects. 

\subsection{Pseudoscalar-pole contribution to HLbL scattering in $(g-2)_{\mu}$ \label{sec:g-2}}

The master formula \Eq{eq:master} requires the knowledge of the TFF at any spacelike virtuality and the dominant contribution comes from the low-energy region with $Q^2\lesssim2~\GeV^2$. Since the single-virtual TFF appears explicitly, this specific kinematic is important. 
For the pion, it was noticed in~\cite{Gerardin:2019vio} that the single-virtual TFF is more difficult to extract on the lattice and this observation motivated the inclusion of a moving frame for the pion to improve the covering of large virtualities. The situation tends to be better for the $\eta$ and $\etap$ mesons due to their heavier masses: we probe larger virtualities and the signal deteriorates more slowly at large virtualities.

A detailed analysis of the weight functions $w_1$ and $w_2$ has been done in \cite{Nyffeler:2016gnb} and we simply mention their most relevant properties. First, the second weight function $w_2$ is typically one order of magnitude smaller than $w_1$ and, at large virtualities, the weight function $w_1$ goes to zero as 
\begin{equation}
w_1(Q_1,Q_2,\tau) \underset{Q_{1}^2\to\infty}{\sim} Q_{1}^{-1} \,, \quad w_1(Q_1,Q_2,\tau)  \underset{Q_{2}^2\to\infty}{\sim}  Q_{2}^{-2} \,, \quad w_1(Q,Q,\tau) \underset{Q^2\to\infty}{\sim} Q^{-2} \,.
\end{equation}
In particular, there is a long tail at constant value of $Q_2^2$. 
Second, the only dependence on the pseudoscalar mass appears as prefactors $(Q_2^2 + m_P^2)^{-1}$ and $( (Q_1+Q_2)^2 + m_P^2)^{-1}$ in $w_1$ and $w_2$ respectively. Thus, we expect the saturation of the integrand, as a function of the $Q_1$ and $Q_2$, to be slower for heavier pseudoscalar mesons. 
In~\cite{Nyffeler:2016gnb}, it was noted that, for the bulk of the contribution, the relation $w_1^{\eta} = (1/6) w_1^{\pi}$ and $w_1^{\etap} = (1/2.5) w_1^{\eta}$ approximately holds. 

\subsubsection{Lattice result} 

\begin{figure}[t]
	\includegraphics*[width=0.32\linewidth]{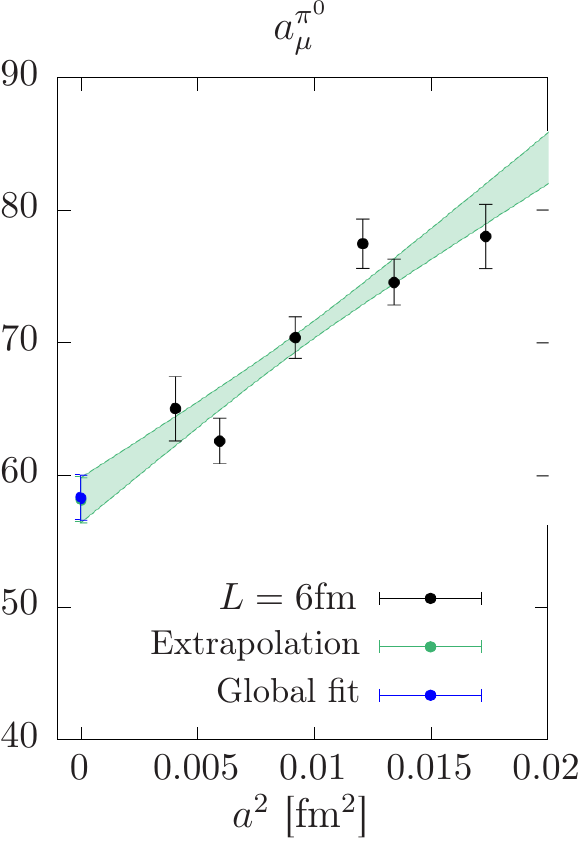}
	\includegraphics*[width=0.32\linewidth]{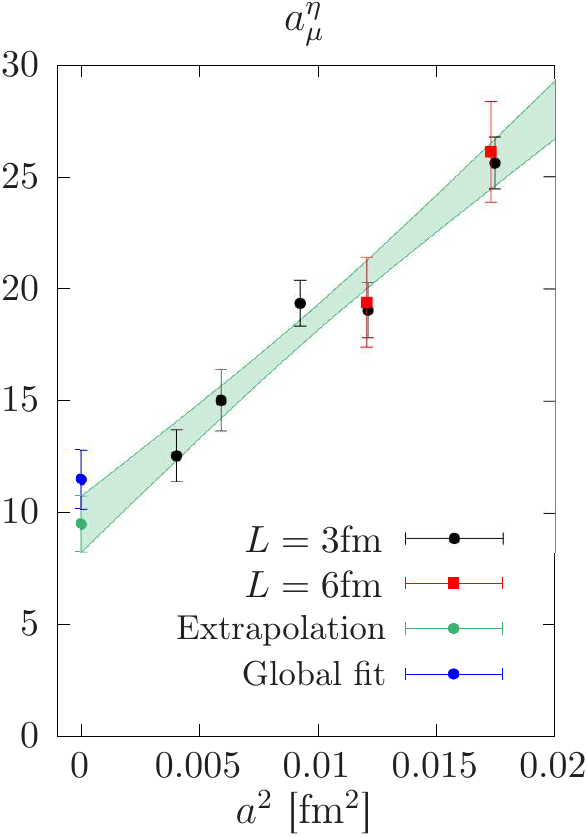}
	\includegraphics*[width=0.32\linewidth]{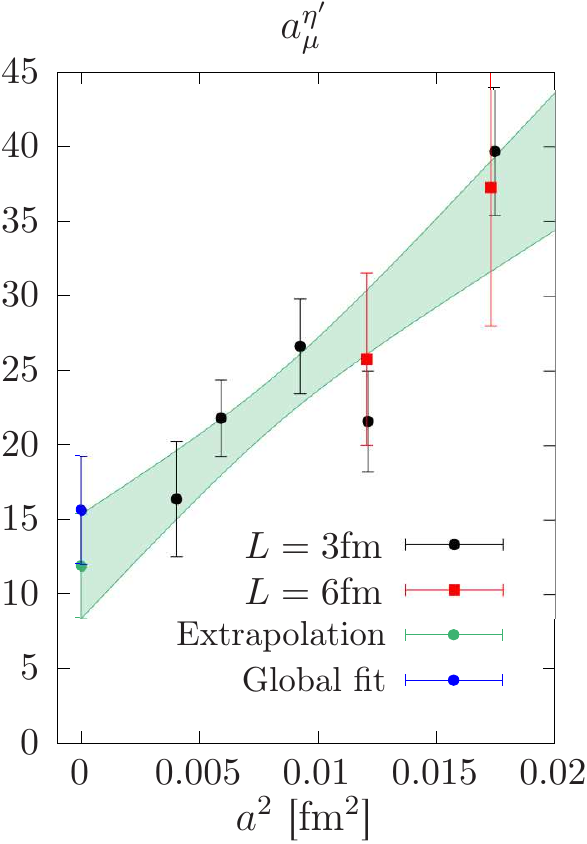}
	\caption{Continuum extrapolation of the pseudoscalar-pole contribution $\amu$ from one typical analysis variation described in \Section{sec:syst}. 
	The black and red points are obtained by fitting the TFFs for a single ensemble to the parametrization of~\Eq{eq:zexp}. 
	For the $\eta$ and $\etap$ mesons, the red points correspond to large-volume ensembles only. We emphasize that single-ensemble fits tend to be instable at large $Q^2$, especially for the pion. Thus the global fit strategy is always our preferred choice. It leads to the blue points at $a=0$.}
\label{fig:amu}
\end{figure}

\begin{figure}[t]
	\includegraphics*[width=0.9\linewidth]{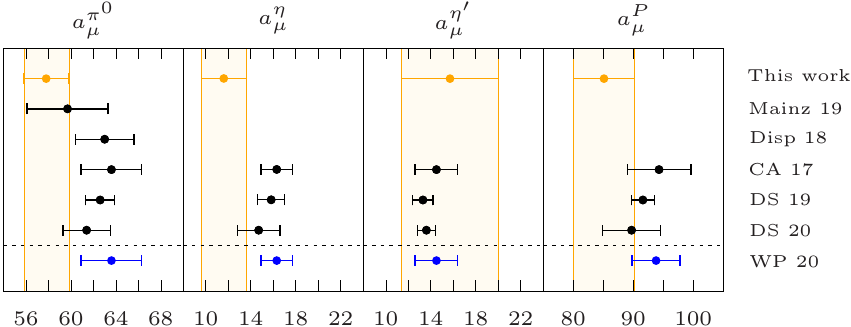}
	\caption{Overview of recent estimates for the pion, $\eta$ and $\etap$-pole contribution to HLbL scattering in muon $g-2$ in units of $10^{-11}$. The data are extracted from the Mainz lattice result~\cite{Gerardin:2019vio}, the data-driven approach (Disp 18)~\cite{Hoferichter:2018dmo}, the Canterbury approximants estimate (CA 17)~\cite{masjuan:2017tvw}, Dyson Schwinger estimates (DS 19)~\cite{Eichmann:2019tjk} and (DS 20)~\cite{Raya:2019dnh}. The last line is the 2020 White Paper estimate~\cite{Aoyama:2020ynm}. }
	\label{fig:cmp}
\end{figure}

We follow two strategies to perform the continuum extrapolation. In the first strategy, we start with computing $\amu$ ensemble per ensemble and then extrapolate the observable to the continuum limit. The result is displayed by the green band in \Fig{fig:amu}. The alternative strategy is to evaluate~\Eq{eq:master} using the parametrization of the TFF in the continuum limit. The result is shown in blue in \Fig{fig:amu}. Both methods lead to compatible results and we quote the second method as our preferred strategy because the fits involved are more stable. It leads to 
\begin{subequations}
\begin{align} 
\amupi &= (57.8 \, \pm \, 1.8_{\rm stat} \, \pm \, 0.9_{\rm syst}) \times 10^{-11}, \\ 
\amueta\, &= (11.6 \, \pm \, 1.6_{\rm stat} \, \pm \, 0.5_{\rm syst} \, \pm \, 1.1_{\rm FSE}) \times 10^{-11}, \\ 
\amuetap &= (15.7 \, \pm \, 3.9_{\rm stat} \, \pm \, 1.1_{\rm syst}\, \pm \, 1.3_{\rm FSE}) \times 10^{-11}.
\end{align} 
\label{eq:sum}
\end{subequations}
Summing all contributions, and taking correlation into account, we obtain the pseudoscalar-pole contribution 
\begin{equation}
\amu = (85.1 \, \pm 4.7_{\rm stat} \, \pm \, 2.3) \times 10^{-11} \,.
\end{equation}
We observe small anticorrelations (with a correlation coefficient of order $-0.1$) between the $\pi^0$ and $\eta$-pole contributions as well as between the $\pi^0$ and $\etap$-pole contributions. The correlation coefficient between the $\eta$ and $\etap$-pole contribution is 0.35.
The pion-pole contribution is in good agreement with previous lattice calculations~\cite{Gerardin:2016cqj,Gerardin:2019vio} that found $\amupi = 59.7(3.6) \times 10^{-11}$. It is compatible at the level of 1.5$\sigma$ with the data-driven estimate~\cite{Hoferichter:2018dmo},  $\amupi = (62.6^{+3.0}_{-2.5}) \times 10^{-11}$ used in the average presented in~\cite{Aoyama:2020ynm}. 
The $\eta$-pole contribution is in slight tension with the White Paper estimate based on Canterbury approximants~\cite{masjuan:2017tvw}, $\amueta = 16.3(1.4) \times 10^{-11}$, but also with the results based of Dyson-Schwinger equations: $\amueta = 15.8(1.2) \times 10^{-11}$~\cite{Eichmann:2019tjk} and $\amueta = 14.7(1.9) \times 10^{-11}$~\cite{Raya:2019dnh}. This tension is mostly due to the momentum region below 0.5~$\GeV$ as indicated by the low value of the decay rate. 
Finally, for the $\etap$-pole, our result is in good agreement with the Canterbury approximant (CA) estimate~\cite{masjuan:2017tvw}: $\amuetap = 14.5(1.9) \times 10^{-11}$ and with the result based on Dyson-Schwinger equations:  $\amuetap = 13.3(0.9) \times 10^{-11}$~\cite{Eichmann:2019tjk} and  $\amuetap = 13.6(0.8) \times 10^{-11}$~\cite{Raya:2019dnh}. 
A summary plot is shown in \Fig{fig:cmp}.

In \Table{tab:contrib} we provide the relative contribution of the pseudoscalar-pole contributions $a_{\mu}^{\rm hlbl,P}$ as a function of the momentum cutoff $Q_{\rm cut}$. Here, the TFF is assumed to be zero if either $|Q_1| > Q_{\rm cut}$ or $|Q_2| > Q_{\rm cut}$. Note that it differs from a cut in the integration range in the master equation~(\ref{eq:master}) as one of the arguments of one of the TFFs in each term of the sum is $Q_1^2 + Q_2^2$. Since the TFF is the central quantity computed in this work, we believe that this choice is more meaningful.
If the contribution from the region $Q^2<0.25~\GeV^2$ represents more than 50\% for the pion, it drops to 31\% for the $\eta$ and even 20\% for the $\etap$. 
Concerning the contribution to the total error, we observe a similar scaling with $Q_{\rm cut}$.
In particular, for the pion, half of the error originates from the region $Q^2<0.25~\GeV^2$ where the signal to noise ratio deteriorates.

\renewcommand{\arraystretch}{1.1}
\begin{table}[t]
\caption{Relative contribution of the pseudoscalar-pole contributions $a_{\mu}^{\rm hlbl,P}$ as a function of the momentum cutoff $Q_{\rm cut}$ (see the main text for explanations). } 
\begin{center}
\begin{tabular}{c@{\hskip 02em}c@{\hskip 02em}c@{\hskip 02em}c}
\hline
$Q^2_{\rm cut} [\GeV^2]$	&	$\pi^0\ (\%)$	&	$\eta\ (\%)$	&	$\etap\ (\%)$	 \\ 
\hline
0.25	&	54	&	31	&	20	  \\ 
0.50	&	71	&	50	&	38	  \\ 
0.75	&	80	&	61	&	51	  \\ 
1.00	&	84	&	68	&	60	  \\ 
1.50	&	89	&	76	&	71	  \\ 
2.00	&	92	&	81	&	78	  \\ 
3.00	&	94	&	86	&	85	  \\ 
5.00	&	97	&	91	&	92	  \\ 
\hline 
\end{tabular}
\label{tab:contrib}
\end{center}
\end{table}

\subsubsection{Discussion of systematic errors \label{sec:systerror}}

\begin{figure}[t]
	\includegraphics*[width=0.3\linewidth]{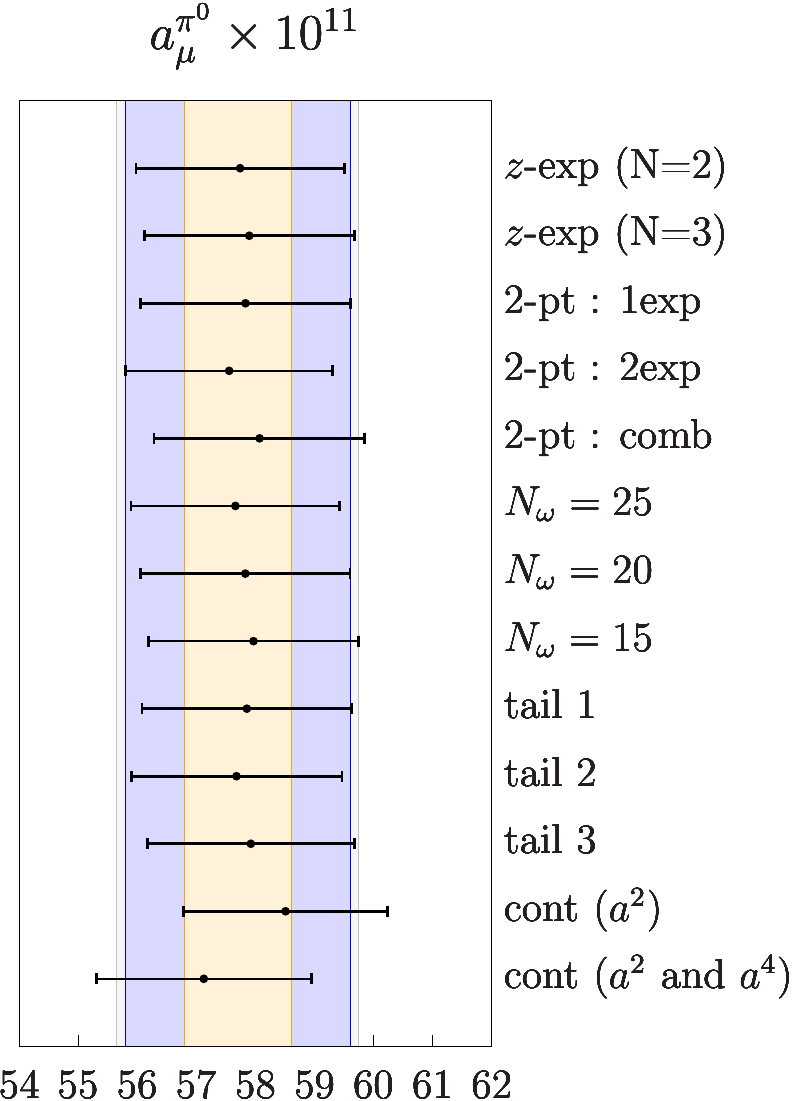} \quad
	\includegraphics*[width=0.3\linewidth]{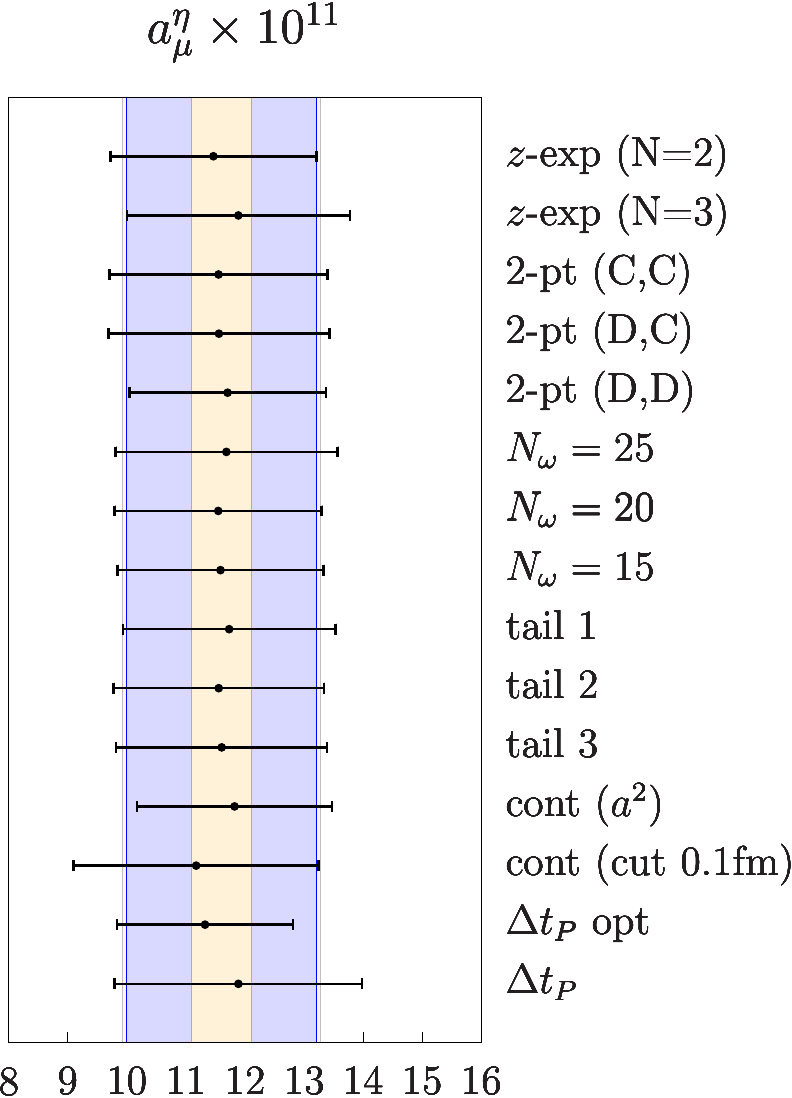} \quad
	\includegraphics*[width=0.3\linewidth]{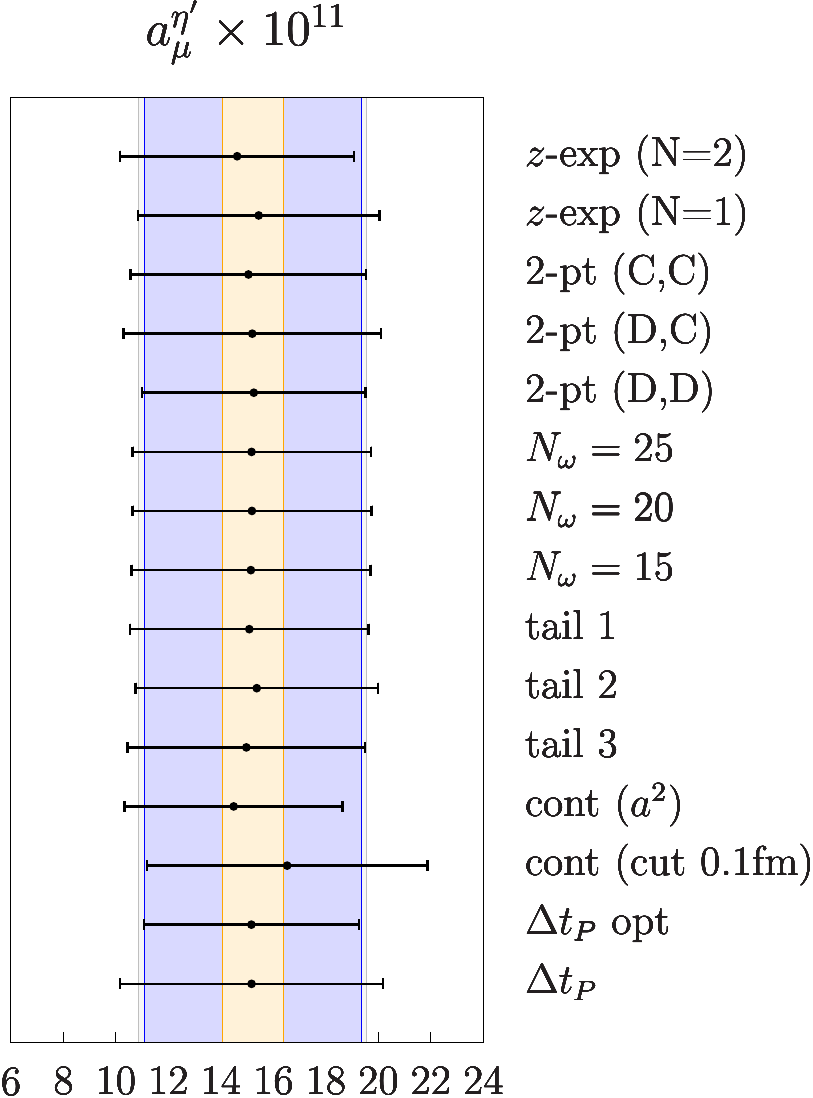}
	\caption{Study of the systematic error on $\amu$ for the pion, $\eta$ and $\etap$. The orange, blue and gray bands correspond to the systematic, statistical and total error respectively. For each point, the result is obtained from a model average over all variations that contain the specific choice. }
	\label{fig:syst}
\end{figure}

In this section, we discuss the different analysis variations that have been included to estimate the systematic errors using the procedure presented in \Section{sec:syst}.  A summary plot is provided in \Fig{fig:syst}.

For the pion, three different procedures have been used to extract the mass and overlap factor with our interpolating operator: we start with a two-exponential fit in a time range that leads to a good chi-square (2exp). In a second step, this fit is used to determine a fit range where the excited state contribution is negligible and the data are fitted by a single-exponential (1exp). Finally, we perform a combined fit where both pion momenta are fitted simultaneously assuming the validity of the dispersion relation (comb). Similarly, for the $\eta^{(\prime)}$ mesons, three different procedures, described in~\cite{spectro}, have been considered. In this case, the masses and overlap factors are obtained by fitting a correlation matrix with both vanishing and nonvanishing pseudoscalar momenta. At vanishing momentum, it is often advantageous to consider the forward derivative $D(t) = C(t+a)-C(t)$ that significantly reduces the statistical noise. Thus, the three variations refer to the combinations $(C,C)$, $(D,C)$ and $(D,D)$ where $(D,C)$ means that the forward derivative $D$ is used at vanishing momenta while the original correlator $C$ is used for the moving frame. 

For all TFFs, three samplings of the orbit shown in \Fig{fig:kin} have been considered. They differ by the number of $\omega$ values for each orbit, $N_{\omega} = 15, 20, 25$. We also considered three methods for the tail correction, that differ by the value of $Q^2_{\rm tail} = (0.75, 0.75, 1.25)~$GeV and/or the model (VMD, LMD, LMD) used to fit the lattice data in the region $Q_1^2,Q_2^2<Q^2_{\rm tail}$. 

For the $\eta$ and $\etap$ TFFs, we also considered two different values of $\Delta t_P$. Our optimal choice, from \Section{sec:topt}, and a more conservative choice.

In the continuum extrapolation of the pion TFFs, we have considered $a^4$ correction as explained in \Section{sec:zexp}. For the $\eta$ and $\etap$ mesons, the lattice data are not precise enough to test higher order terms in the continuum extrapolation. Instead, we have performed cuts in the lattice spacing by removing all points with $a>0.1~$fm.

The impact of each particular choice is shown in \Fig{fig:syst} where each point is obtained by averaging over all variations that are compatible with this choice. Our total error is dominated by statistics, especially for the $\eta$ and $\etap$ mesons.

\section{Conclusion \label{sec:ccl}} 
		
In this paper, we have presented a first abinitio calculation of the $\eta$ and $\etap$ transition form factors at the physical point, in the kinematical range relevant for computing the HLbL contribution to the anomalous magnetic moment of the muon. Results for the pion TFF are also provided. 
As compared to the pion TFF, the extraction of the $\eta$ and $\etap$ TFFs requires a proper treatment of the mixing in the $\eta-\etap$ system~\cite{spectro}.
Thanks to the implementation of advanced numerical methods, we have been able to compute and get a signal for all disconnected contributions: the diagram with a single pseudoscalar loop is by far dominant, the other contributions are SU(3)$_f$ suppressed and contribute at the percent level.
 
Our result for the pion TFF is in good agreement with previous lattice calculations~\cite{Gerardin:2016cqj,Gerardin:2019vio} and with experimental data in the single-virtual regime. 
Concerning the $\eta$ TFF, our result turns out to be in good agreement with previous estimates based on Canterbury approximants~\cite{masjuan:2017tvw} or Dyson-Schwinger equations~\cite{Eichmann:2019tjk,Raya:2019dnh} for virtualities above $0.4~\GeV^2$. 
However, we find a slight tension, at the level of 1.7$\sigma$, for the normalization of the $\eta$ TFF. 
In the current setup, extracting the normalization of the TFF is challenging as it requires an extrapolation with data becoming noisier as we approach vanishing virtualities. Additional larger-volume ensembles could further improve the result.
The single-virtual $\eta$ form factor is also compatible with experimental measurements by CELLO~\cite{CELLO:1990klc} and CLEO~\cite{CLEO:1997fho}, with a slight tension for the lowest CELLO bin.
Finally, the result for the $\etap$ TFF is in good agreement with other approaches in the double-virtual regime. In the single-virtual regime, our result is compatible (at the level of 1-2$\sigma$) with existing measurements from CELLO~\cite{CELLO:1990klc}, CLEO~\cite{CLEO:1997fho},  BaBar~\cite{BaBar:2018zpn,BaBar:2011nrp} and the L3 experiment at LEP~\cite{L3:1997ocz}. We observe a slight tension with the results based on the Dyson-Schwinger equation~\cite{Eichmann:2019tjk}.
  
Using our parametrization of the TFFs at the physical point, we provide an abinitio estimate of the light-pseudoscalar meson-pole contribution to HLbL.
For the pion, we confirm the previous lattice calculations with Wilson-clover quarks~\cite{Gerardin:2019vio} where a good agreement was observed with the data-driven estimate of Ref.~\cite{Hoferichter:2018dmo}. 
The extraction of the $\eta$ and $\etap$ pole contributions is much more challenging but a precision of 17\% and 27\% has been achieved.
Previous estimates of the $\eta^{(')}$ pseudoscalar-pole were based on Dyson-Schwinger equations~\cite{Eichmann:2019tjk,Raya:2019dnh} or Canterbury approximants~\cite{masjuan:2017tvw}, relying on the availability of experimental data. For the $\eta$ meson we find a slightly smaller value, although compatible at the level of 1.9 combined standard deviations. This can be explained by our lower value for the normalization of the form factor as compared to the experimental value quoted in~\cite{ParticleDataGroup:2018ovx}. 

Summing all contributions in~\Eq{eq:sum}, we obtain the main result of this paper
\begin{equation} 
\amu = (85.1 \, \pm 4.7_{\rm stat} \, \pm \, 2.3_{\rm syst}) \times 10^{-11} \,.
\end{equation}
This value is compatible with the White-Paper estimate, $\amu=93.8^{+4.0}_{-3.6} \times 10^{-11}$~\cite{Aoyama:2020ynm} at the level of 1.4 combined standard deviations, and with a competitive uncertainty. 

In the future, adding large-volume ensembles at smaller lattice spacings for the $\eta$ and $\etap$ TFFs would help to better constrain the normalization of the form factor and to clarify the tension on the $\eta \to \gamma \gamma$ decay rate. The main advantage of large physical volume is to provide a better sampling of the low-virtuality region close to the origin. As shown on \Table{tab:contrib}, reducing the error on the decay rate has a strong impact on the $\pi^0$ and $\eta$-pole contributions and, to a lesser extent, on the $\etap$-pole contribution. Another possibility would be to use the framework presented in~\cite{Meng:2021ecs,Feng:2018qpx} to directly compute the normalization of the TFF.


\begin{acknowledgments}
We thank all the members of the Budapest-Marseille-Wuppertal collaboration for helpful discussions and the access to the gauge ensembles used in this work. 
This publication received funding from the Excellence Initiative of Aix-Marseille University -- A*Midex, a French ``Investissements d'Avenir" programme, Grant No. AMX-18-ACE-005 and from the French National Research Agency under Grant No. ANR-20-CE31-0016. 
The computations were performed on Joliot-Curie at CEA's TGCC, on Jean Zay at IDRIS, on SuperMUC-NG at Leibniz Supercomputing Centre in M\"unchen, on HAWK at the High Performance Computing Center in Stuttgart and on JUWELS at Forschungszentrum J\"ulich. We thank GENCI (Grants No. A0080511504, No. A0100511504 and No. A0120511504) and the Gauss Centre for Supercomputing (projects pn73xi and wprecision) for awarding us computer time on these machines.
Centre de Calcul Intensif d'Aix-Marseille (CCIAM) is acknowledged for granting access to its high performance computing resources.
\end{acknowledgments}

\appendix

\section{Coefficients of the $z$-expansion}  \label{app:tables}

In this appendix we provide the coefficients of the $z$-expansion (\Table{tab:coeff_eta}) and the associated correlation matrix as described in \Section{sec:zexp}. The results correspond to one of the many analysis variations described in \Section{sec:syst} and the errors are statistical only.
 
\renewcommand{\arraystretch}{1.2}
\begin{table}[h]
\caption{Coefficients of the $z$-expansion at the physical point for the pion, $\eta$ and $\etap$ mesons with $N=2$. We also quote the value of $Q_{\max}^2$ and $\Lambda$ that enters in \Eq{eq:zexp}.} 
\begin{center}
\begin{tabular}{ccccccccccccc}
\hline
Meson	&	$c_{00}$	&	$c_{01}$	&	$c_{11}$	&	$c_{20}$	&	$c_{21}$	&	$c_{22}$	& 	$Q_{\max}^2$	&  $\Lambda$ \\ 
\hline
$\pi^0$	& $+0.2372(22)$ &	$-0.064(0
5)$	&	$-0.26(11)$	&	$+0.082(55)$	&	$+0.040(65)$ 	&	$-0.060(53)$ 	&	2.2 GeV$^2$ & 775 MeV	 \\ 
$\eta$	& $+0.198(11)$ &	$-0.021(15)$	&	$-0.39(15)$	&	$+0.154(94)$	&	$-0.22(19)$	&	$+0.37(47)$	&	5 GeV$^2$ & 775 MeV	 \\ 
$\etap$	& $+0.296(31)$   &	$-0.055(25)$	&	$-0.29(16)$	&	$+0.001(103)$ 	&	$+0.04(26)$	&	$-0.01(71)$	&	5 GeV$^2$ & 1 GeV	 \\ 
\hline 
\end{tabular}
\label{tab:coeff_eta}
\end{center}
\end{table}
 
\begin{equation}
\mathrm{cor}(c_{nm})_{\pi^0} = \begin{pmatrix}
 +1.00 & -0.51 & +0.20 & -0.26 & +0.37 & -0.27 \\
-0.51 & +1.00 & -0.42 & +0.40 & -0.61 & +0.45  \\
+0.20 & -0.42 & +1.00 & -0.97 & +0.70 & +0.20 \\
-0.26 & +0.40 & -0.97 & +1.00 & -0.81 & -0.10  \\
+0.37 & -0.61 & +0.70 & -0.81 & +1.00 & -0.43 \\
-0.27 & +0.45 & +0.20 & -0.10 & -0.43  &+1.00 \\
\end{pmatrix} \
\label{eq:cov_pi} 
\end{equation}

\begin{equation}
\mathrm{cor}(c_{nm})_{\eta} = \begin{pmatrix}
+1.00 & +0.08 & +0.37 & -0.42 & -0.04 & +0.19 \\
+0.08 & +1.00 & -0.25 & +0.23 & -0.74 & +0.36 \\
+0.37 & -0.25 & +1.00 & -0.89 & +0.40 & -0.02  \\
-0.42 & +0.23 & -0.89 & +1.00 & -0.44 & -0.21  \\
-0.04 & -0.74 & +0.40 & -0.44 & +1.00 & -0.60  \\
+0.19 & +0.36 & -0.02 & -0.21 & -0.60 & +1.00 \\
\end{pmatrix} 
\label{eq:cov_eta} 
\end{equation}

\begin{equation}
\mathrm{cor}(c_{nm})_{\etap} = \begin{pmatrix}
+1.00 & -0.06 & +0.01 & -0.20 & -0.33 & +0.21 \\
-0.06 & +1.00 & -0.51 & +0.36 & -0.63 & +0.17 \\
+0.01 & -0.51 & +1.00 & -0.78 & +0.64 & -0.22 \\
-0.20 & +0.36 & -0.78 & +1.00 & -0.38 & -0.29  \\
-0.33 & -0.63 & +0.64 & -0.38 & +1.00 & -0.54  \\
+0.21 & +0.17 & -0.22 & -0.29 & -0.54 & +1.00 \\
\end{pmatrix} 
\label{eq:cov_etap} 
\end{equation}

\section{Disconnected contributions : estimator for the vector loops}  \label{sec:discV}

In this appendix we compare different estimators to evaluate the vector loops that appear in the second and fourth disconnected diagrams of \Fig{fig:wick}. We also provide estimates of those subleading diagrams to the pion-pole contribution to $\ahlbl$. 

\subsection{Noise reduction techniques}

In the isospin limit, we are interested in the vector loop function for the light-minus-strange contribution
\begin{align}
L^{(l-s)}_{V;\mu}(x_0;\qv) &= \sum_{\vec{x}} \Tr\left[ \gamma_{\mu} D^{-1}_l(x,x)  - \gamma_{\mu} D^{-1}_s(x,x)  \right] e^{i \qv \cdot \xv}  \\
&= (m_s - m_l) \sum_{\vec{x}} \Tr\left[ \gamma_{\mu} D^{-1}_l D^{-1}_s(x,x) \right] e^{i \qv \cdot \xv}
\end{align}
where $D_f$ denotes the Dirac operator for a fermion of mass $m_f$.  In this appendix we use Wilson-like fermion notations. 
A standard estimator is to use the same set of $N$ volume stochastic sources for both light and strange contributions~\cite{Gulpers:2014jaq} 
\begin{equation}
L^{(l-s)}_{V;\mu}(x_0;\qv) = (m_s-m_l) \frac{1}{N} \sum_{n=1}^{N} \sum_{\vec{x}} \left[ \eta_n^{\dag}(x) \gamma_{\mu} D^{-1}_l D^{-1}_s \eta_n(x) \right] e^{i \qv \cdot \xv}
\end{equation}
where the volume sources satisfy $\langle \eta(x) \eta^{\dag}(y) \rangle = \delta_{x,y}$. This estimator is noisy and typically thousands of inversions are required to reach the gauge noise. 
In~\cite{Giusti:2019kff}, the authors introduced the split-even stochastic estimator
\begin{equation}
L^{(l-s)}_{V;\mu}(x_0; \qv) = (m_s-m_l) \frac{1}{N} \sum_{n=1}^{N} \sum_{\vec{x}} \left[ \left( \eta_n^{\dag} D_l^{-1} \right)(x) \gamma_{\mu} \left( D^{-1}_s \eta_n \right)(x) \right] e^{i \qv \cdot \xv}
\end{equation}
which drastically reduces the variance of the estimator for a comparable numerical cost. Finally, both estimators can be combined with LMA where the low-mode contribution is computed exactly, without adding stochastic noise. When the low modes of the Dirac operator are used to deflate the lattice Dirac operator, the overhead of LMA is small.

\begin{figure}[t]
	\includegraphics*[width=0.65\linewidth]{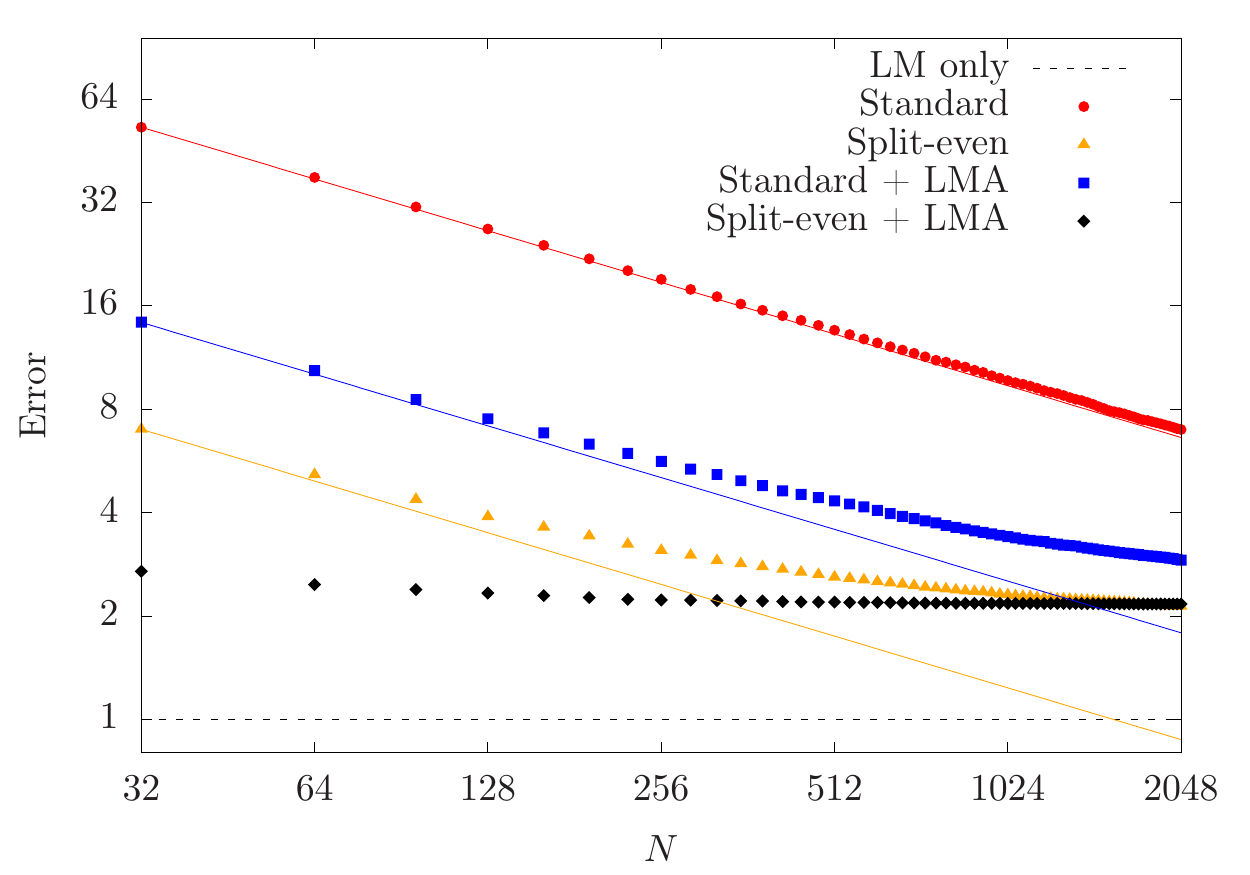}
	\caption{Square root of the variance of the vector loop function for the four estimators described in the text. The lines represent a perfect scaling of the variance assuming the variance is dominated by the stochastic noise. The horizontal dashed line is the error from the low-mode contribution only and is used as a normalization scale.}
	\label{fig:gn}
\end{figure}

In this appendix, we compare the performances of the four estimators discussed above.
We use the variance of the loop function as a benchmark of the estimators. 
Since the error on the loop function is constant in time, we average the variance over all time slices and all directions $\mu=0,1,2,3$.  
The results are obtained on the $L=6~$fm ensemble at a lattice spacing of $a=0.13~$fm and with $\qv=\vec{0}$. We use 1000 eigenvectors of the even-odd Dirac operator (i.e. 2000 eigenvectors of the Dirac operator) to compute the low-mode contribution. The results are depicted in \Fig{fig:gn}.

We see that the split-even estimator outperforms the standard estimator, even when the latter is combined with LMA. Less than 1000 inversions are needed to reach the gauge noise. When the split-even estimator is combined with LMA, only $O(10)$ sources are needed to reach the gauge noise. Combining LMA with the split-even stochastic estimator, the computer time spent to compute those disconnected loops is completely subdominant.

\subsection{Contribution of the subleading diagrams to $\amu$}

\begin{figure}[t]
	\includegraphics*[width=0.4\linewidth]{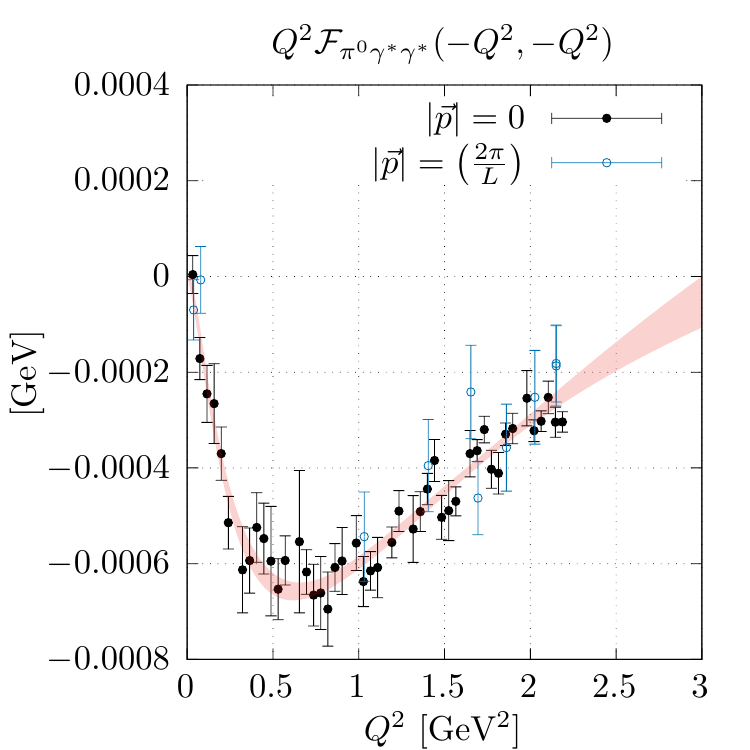}
	\includegraphics*[width=0.4\linewidth]{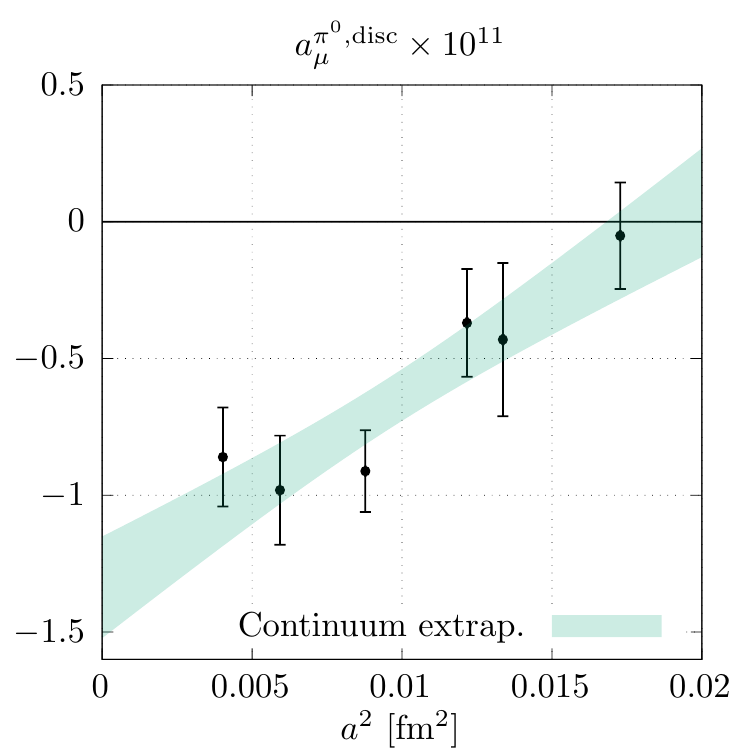}
	  
	\caption{Contribution of disconnected diagram with a single vector loop to the pion transition form factor at our finest lattice spacing (left panel) and to $\amupi$ in the continuum limit (right panel).}
	\label{fig:disc1}
\end{figure}

The second diagram in \Fig{fig:wick}, with a single vector loop, is the only disconnected contribution to the pion TFF. We emphasize that this contribution, which vanishes in the SU(3) flavor limit, is included in the main analysis. The goal of this section is to isolate its contribution to $\amupi$. 
We start with the decomposition $\FFpi = \FFpi^{\rm conn} + \FFpi^{\rm disc}$ and treat the disconnected contribution as a small correction. The result for $\FFpi^{\rm disc}$ on one ensemble is  shown on left panel of \Fig{fig:disc1} where the data is fitted using the $z$-expansion described in \Section{sec:zexp} but setting $P=1$. We observe that this contribution decreases much faster than the connected contribution at large virtualities. Finally, the contribution to $\amupi$ is obtained from a linearized version of the master formulas \Eq{eq:master} and the results are shown in the right panel of \Fig{fig:disc1}. In the continuum limit, and using a linear fit in $a^2$ we find $-1.33(19) \times 10^{-11}$ compatible with the published result $-1.0(0.5) \times 10^{-11}$ from \cite{Gerardin:2019vio}, but with much smaller uncertainties. Although this contribution is small, it cannot be neglected for the pion at our level of precision.

The last diagram in \Fig{fig:wick} contains a pseudoscalar loop and contributes only to the $\eta$ and $\etap$ TFFs. The relative contribution is shown in \Fig{fig:C}. For those states, the two contributions that involves a single vector loop partly compensate each other and their contribution turns out to be negligible at the current level of precision.

\bibliography{biblio}{}

\end{document}